\documentclass[10pt, journal, twocolumn]{IEEEtran} %

\usepackage{cite}       

\usepackage{graphicx, ulem}   

\usepackage{psfrag}    

\usepackage{subfigure} 

\usepackage{url}       

\usepackage{stfloats}  

\usepackage{amsmath}   
\usepackage{array}
 \makeatletter
 \let\NAT@parse\undefined
 \makeatother
\usepackage{times}
\usepackage{epsfig,graphicx,color,pstricks}
\usepackage{amsmath,amssymb,amsthm,amsbsy,amsfonts,latexsym}
\usepackage{bm}
\usepackage{psfrag}
\usepackage[mathscr]{eucal}
\usepackage{cite}
\usepackage[all]{xy}
\usepackage{subfigure}
\usepackage{graphics}

\newtheorem{thm}{Theorem}
\newtheorem{cor}[thm]{Corollary}

\theoremstyle{remark}
\newtheorem{rem}{Remark}
\theoremstyle{definition}

\pdfoutput=1

\begin{document}

\title{Multi-user Cognitive Interference Channels:\\
 A Survey and New Capacity Results}

\author{
\IEEEauthorblockN{Diana Maamari, Daniela Tuninetti and Natasha Devroye}
\thanks{%
This work was performed while the authors were 
with the Electrical and Computer Engineering Department of the University of Illinois at Chicago (UIC), Chicago, IL 60607 USA. 
Diana Maamari is now with Huawei Technologies, Rolling Meadows, IL USA (e-mail: Diana.Maamari@huawei.com). 
Natasha Devroye and Daniela Tuninetti are with UIC (e-mail: \{devroye, danielat\}@ uic.edu).
Their work at UIC was partially funded by NSF under award number 0643954 and 1017436; the contents of this article are solely the responsibility of the authors and do not necessarily represent the official views of the NSF.
}}

\maketitle

\begin{abstract}
This paper provides a survey of the state-of-the-art information theoretic analysis for {\it overlay multi-user (more than two pairs) cognitive networks} and reports new capacity results. In an overlay scenario, cognitive / secondary users share the same frequency band with licensed / primary users to efficiently exploit the spectrum. They do so without degrading the performance of the incumbent users, and may possibly even aid in transmitting their messages as cognitive users are assumed to possess the message(s) of primary user(s) and possibly other cognitive user(s). 
The survey begins with a short overview of  
the two-user overlay cognitive interference channel. The evolution from two-user to three-user overlay cognitive interference channels is described next, followed by generalizations to multi-user (arbitrary number of users) cognitive networks. 
The rest of the paper considers $K$-user cognitive interference channels  
with different message knowledge structures at the transmitters. 

Novel capacity inner and outer bounds are proposed.
Channel conditions under which the bounds meet, thus characterizing the information theoretic capacity of the channel, 
for both Linear Deterministic and Gaussian channel models, are derived. 

The results show that for certain channel conditions {\it distributed cognition}, or having a cumulative message knowledge structure at the nodes, may not be worth the overhead as (approximately) the same capacity can be achieved by having only one {\it global cognitive user} whose role is to manage all the interference in the network. The paper concludes with future research directions.
\end{abstract}

\section{Introduction}
\label{sec:intro}

More efficient usage of the spectrum is needed given the ever increasing demand for wireless broadband services. Cognitive radio combined with spectrum sharing have been proposed as a solution to this apparent spectrum crunch.  
This would involve smart new ``cognitive'' wireless devices intelligently coexisting with users with priority access to the spectrum, either minimally impacting them, or not impacting them at all.

The usage of cognitive radios allows for ``cognition'' in wireless networks, which broadly speaking implies that the wireless devices are able to adapt in real-time to the wireless environment. This may translate to a number of assumptions and/or schemes technically.  We list three common sets of assumptions and their corresponding nomenclature below~\cite{goldsmith2009breaking}. 

\begin{enumerate}

\item
The approach where cognitive radios sense white spaces (time, space or frequency voids) and adjust their transmissions to fill the sensed voids has been referred to as {\it interweave} and  avoids interference altogether, at the (possible) expense of spectral efficiency. 

\item
Contrary to keeping its transmission orthogonal to the primary user's transmissions, as is the case in the interweave paradigm, the  
{\it underlay} paradigm allows a cognitive radio to simultaneously transmit with primary user(s), if the interference caused at the primary receiver(s) is kept below a certain threshold that is commonly referred to as the ``interference temperature constraint.'' In this case, a cognitive radio adjusts its transmission power in order to satisfy the interference temperature. The maximal tolerable interference for the surrounding users, as well as channel state information of the interfering channel gains, 

is assumed available at the cognitive transmitter.

\item
The case where the cognitive devices have additional information of codebooks, messages or channel gains of other user(s), and they  simultaneously transmit with primary license holders, is referred to as an 
{\it overlay} paradigm, or a Cognitive Interference Channel (CIFC)~\cite{goldsmith2009breaking}. 

\end{enumerate}

{\it 
In this paper we focus on an overlay form of cognition, where secondary users have a-priori non-causal message knowledge of primary license holder(s). All nodes furthermore are assumed to have global codebook and channel state information, as is common in an information theoretic analysis of complex network models.
} 
Intuitively, this idealized assumption allows the cognitive radios to cooperate in sending the  primary user's message while at the same time transmitting their own message by using an interference mitigation technique.

In this paper, we survey the fundamental limits of communication for a multi-user overlay cognitive networks with an {\it arbitrary} number of secondary / cognitive user(s) having non-causal message knowledge of primary user(s). The users transmit in the same frequency band and thus in general interfere with one another. The performance metric considered is the information theoretic notion of channel capacity. In other words, we are interested in the maximum rate of communication for which arbitrarily small probability of error can be achieved by every user, which may be seen as a benchmark when building practical systems. We will focus on results for general memoryless  and practically relevant additive white Gaussian noise (AWGN)~\cite{Shannon:1948} channel models, as well as high signal to noise ratio (SNR) approximations of Gaussian channels called linear deterministic channels (LDA)~\cite{AVHESTIMERPHD}.

\subsection{Motivation for Non-Causal Message Knowledge}

The asymmetric  form of cooperation of an overlay cognitive multi-user network,  in which cognitive users possess primary users' messages prior to transmission but not vice versa,  may be motivated in a number of ways. For example, if certain receivers were not able to decode their own messages (due to for example packets being lost or damaged), they may request a re-transmission. If other transmitters were able to hear and decode the original transmission, during the re-transmission phase, then knowledge of messages at other transmitters is justified. 
It may also be justified as an upper bound to what a real cognitive transmitter may be able to do, under the assumption that it possesses the primary's codebook, and hence listens and tries to decode the primary users' messages. 

The problem of Coordinated Multipoint (CoMP) joint transmission, also known as base station cooperation, has been considered at length over the decades.  Such models usually consider a  network with base stations  connected via unrestricted backhaul links (error-free and unlimited
capacity) over which messages or subsets of messages can be shared.  The analytically tractable {\it Wyner} model~\cite{A_Wyner} (where nodes are placed on a line and suffer from interference only from a limited number of left and right neighbors) has been a widely adopted model for studying the advantages of base station cooperation in the downlink of cellular networks. In~\cite{Lozano_fun_limits} the authors show that achievable rates with cooperation are upper bounded by a theoretic limit that is independent of the transmit power; thereby proving that arbitrarily high gains can not be achieved through cooperation. The work in~\cite{Multicell_MIMO_Shlomo} provides an overview of the information theoretic results and techniques to study multi-cell MIMO cooperation.  An overlay multi-user cognitive interference network is a form of CoMP or network with base station cooperation in which there are only backhaul links in certain directions, leading to an asymmetry in the message knowledge structure.

\subsection{Contributions} 

In this paper we give an overview of the state-of-the-art results for overlay cognitive networks dating back to the two-user cognitive interference channel ($2$-CIFC) and its reported capacity results. The survey continues with the introduction of the three-user extension of the $2$-CIFC. For the $3$-CIFC different models of cognition, or different message knowledge structure at the transmitters, are possible. 
After overviewing known results for the $3$-CIFC, 

the results when the number of transmitters in the network is an arbitrary integer greater than three are then presented.

Our major contributions include:
\begin{enumerate}

\item
The derivation of new outer bounds on the capacity regions of $K$-CIFC.

\item
Sum-capacity achieving schemes for both the LDA and the Gaussian noise symmetric 
$K$-CIFC are presented. 

\item
Interestingly, we show that under certain symmetric channel conditions, the throughput / sum-capacity of a multi-user cognitive interference with a {\it cumulative} message knowledge structure (or distributed cognition) can be achieved with a much simpler message knowledge structure. In particular, we show that at high SNR, having an interference channel with a cognitive relay is throughput / sum-capacity achieving; the same holds at finite SNR but to within a finite constant additive gap.

\item 
The asymmetric $K$-CIFC is considered and capacity results for the LDA are presented. Translation of these results at finite SNR for the Gaussian noise model is part of on-going work.

\end{enumerate}

The general information theoretic study of cognitive networks is extensive and we do not attempt to survey it all; we focus on genie-aided multi-user cooperative networks in the sense that we assume messages are known to secondary users prior to transmission. 
The interference channel with partial transmitter cooperation, the causal cognitive interference channel (channels with cooperation between transmitters) considered in~\cite{Cardone_IT} and~\cite{Cardone_JSAC}, the cognitive interference channel with fading~\cite{Diana_Ergodic_TWC} are beyond the scope of this paper.

\subsection{Paper Organization}

The paper is organized as follows.
Section~\ref{sec:ChModel} introduces notation, definitions, and channel models;
Section~\ref{sec:survey} briefly surveys known results for multi-user cognitive interference channels;
Section~\ref{sec:NEWouter} derives novel outer bounds, which are then matched to novel achievable schemes
for the linear deterministic channel in Section~\ref{sec:NEWlda}, and for the Gaussian noise channel in
Section~\ref{sec:NEWawgn}.
Section~\ref{sec:conc} concludes the paper.
Some proofs may be found in Appendix.

This survey is meant to offer an entry point on the literature on the subject of cognitive networks to readers who are familiar with communication theory and point-to-point information theory~\cite{Cover:InfoTheory}, but not necessarily with the latest advances in network information theory~\cite{ElGamalKim:InfoTheory}. In order to make its content accessible to a wide audience, we have provided comments, insights and references to the most technical aspects of the discussion.

\begin{figure*}
\centering
\includegraphics[width=\textwidth]{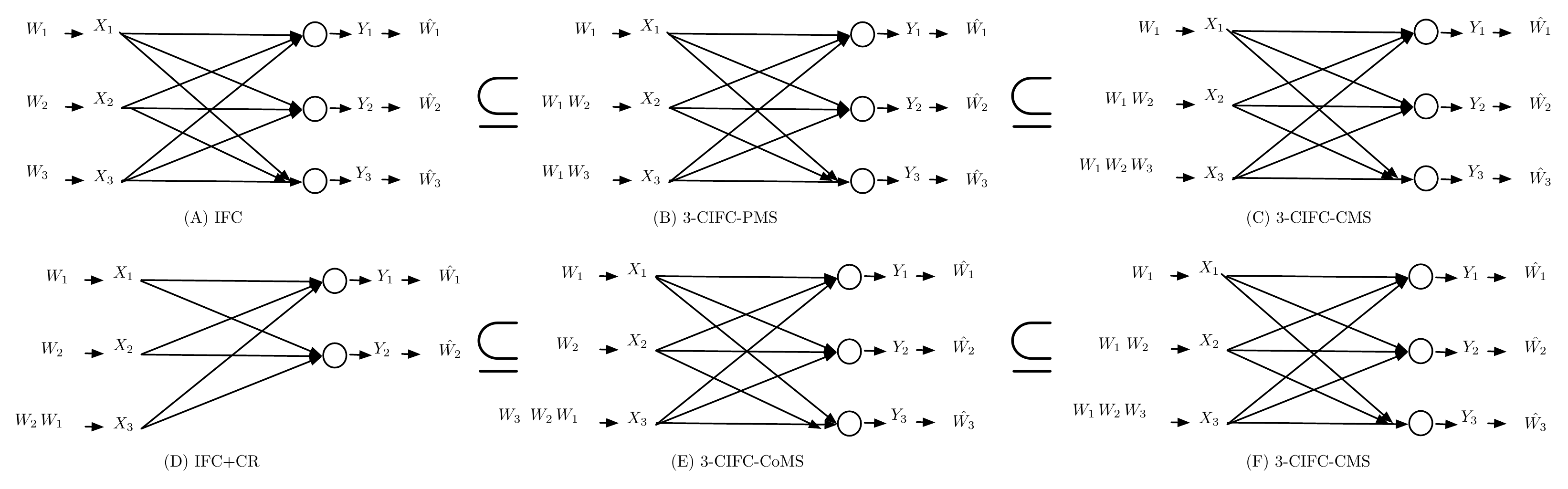}
\caption{$K$-user cognitive interference channels with different message knowledge structures at the transmitters.} \label{fig:PMSinCMS}
\end{figure*}

\section{Channel Models} 
\label{sec:ChModel}

Before discussing known and new results, we introduce the formal information theoretic definition of the problem. We start with describing our notation convention.

\subsection{Notation}
In the following we shall follow the notation convention of~\cite{ElGamalKim:InfoTheory}.
Lower case variables are instances of upper case random variables which take on values in calligraphic alphabets.
The set of integers from $n_1$ to $n_2$ is denoted by $[n_1 : n_2]$.
$Y^{j}$ is a vector of length $j$ with components $(Y_1,\ldots,Y_j)$.
For a vector $X$ and an index set $\mathcal{I}$, $X_{\mathcal{I}} = (X_i : i\in \mathcal{I})$.
$\mathcal{N}(\mu,\sigma^2)$ denotes the density of a complex-valued circularly symmetric Gaussian random variable with mean $\mu$ and variance $\sigma^2$.
$P(.)$ denotes a probability distribution function (a subscript indicating the random variable may be included to avoid confusion), 
$\mathbb{P}[.]$ the probability measure (for the probability of an event),
and $\mathbb{E}[.]$ the expectation. 
The mutual information between random variables $X$ and $Y$ is denoted by $I(X,Y) = \mathbb{E}\left[\log\frac{P(X,Y)}{P(X)P(Y)}\right]$ where $P(X)$ and $P(Y)$ are the marginal distributions of $P(X,Y)$.

\subsection{The General Memoryless CIFC}
\label{sec:ChModel:general:CIFC-CoMS and CMS}

\paragraph*{CIFC-CoMS}
The general memoryless $K$-user Cognitive InterFerence Channel with Cognitive only Message Sharing ($K$-CIFC-CoMS) is shown in Fig.~\ref{fig:PMSinCMS}(E). It consists of $K$ source-destination pairs sharing the same physical channel, where one transmitter has non-causal knowledge of the messages of all the other transmitters. 
Here transmitters $[1:K-1]$ are referred to as primary users and are assumed to have no cognitive abilities. Transmitter~$K$ is non-causally cognizant of the messages of the primary users. More formally, the $K$-CIFC-CoMS channel consists of:
\begin{itemize}
\item 
channel  inputs $X_i \in \mathcal{X}_i, \ i\in[1:K]$, 
\item 
channel outputs $Y_i \in \mathcal{Y}_i, \ i\in[1:K]$, and
\item 
a memoryless channel with joint transition probability distribution (or conditional channel distribution) ${P}(Y_1,\ldots,Y_K|X_1,\ldots,X_K)$.

\end{itemize}

A code with non-negative rate vector $(R_1,\ldots,R_K)$ 
is defined by:
\begin{itemize} 
\item 
Mutually independent messages $W_i$, one for each transmitter $i\in[1:K]$, that are uniformly distributed over $[1:2^{N R_i}]$, where $N$ denotes the block length and $R_i$ the rate in bits per channel use.
\item 
Encoding functions $f_i^{(N)} : 
[1:2^{N R_i}] \to \mathcal{X}_i^N$ such that $X_i^N := f_i^{(N)}(W_i)$, for primary user $i\in[1:K-1]$, while 
$f_K^{(N)} : [1:2^{N R_1}] \times \ldots \times [1:2^{N R_K}] \to \mathcal{X}_K^N$ such that
$X_K^N := f_K^{(N)}(W_1,\ldots,W_K)$ for the cognitive user. 
\item Decoding functions $g_i^{(N)} : \mathcal{Y}_i^N \to [1:2^{N R_i}]$ such that $\widehat{W}_i = g_i^{(N)}(Y_i^N)$, $i\in[1:K]$.
\item 
The (average over all messages) probability of error for user $i\in[1:K]$ is denoted as $\mathbb{P}[\widehat{W}_i \not= W_i]$.
\end{itemize}
The capacity region of the $K$-CIFC-CoMS consists of all  
rate tuples $(R_1,\ldots,R_K)$ for which there exists a sequence of codes indexed by the block length $N$ such that the probability of error of every user can be made arbitrary small, formally, such that $P_{e}^{(N)} := \max_{i\in[1:K]} \mathbb{P}[\widehat{W}_i \not= W_i]\rightarrow 0$ as $N \rightarrow \infty$.%
\footnote {This specific way of writing the overall system error probability $P_{e}^{(N)}$ (as the maximum of individual error probabilities) highlights that the capacity of an IFC without node cooperation, similarly to the broadcast channel~\cite[Chapters~5 and~8]{ElGamalKim:InfoTheory}, does not depend on the joint channel conditional distribution ${P}(Y_1,\ldots,Y_K|X_1,\ldots,X_K)$ but only on the marginal
distributions ${P}(Y_i|X_1,\ldots,X_K), \ i\in[1:K]$---a fact that may be leveraged in deriving outer bounds. }
Note that the capacity under average probability of error criteria $\mathbb{P}[\widehat{W}_i \not= W_i] = \sum_{k} \mathbb{P}[W_i=k] \mathbb{P}[\widehat{W}_i \not=k | W_i=k], \ i\in[1:K],$ may be larger than the capacity under maximal probability of error criteria $\max_k \mathbb{P}[\widehat{W}_i \not=k | W_i=k], \ i\in[1:K],$~\cite[Chapter~4]{ElGamalKim:InfoTheory}.

\bigskip
\paragraph*{CIFC-CMS}
The general memoryless $K$-user Cognitive InterFerence Channel with Cumulative Message Sharing ($K$-CIFC-CMS) is shown in Fig.~\ref{fig:PMSinCMS}(F). It consists of $K$ source-destination pairs sharing the same physical channel, where there are $K-1$ cognitive transmitters and one primary user. 
Here transmitters $i\in [2:K]$ are referred to as cognitive users and are assumed to have non-causal message knowledge of the users' messages with lesser index. The $K$-CIFC-CMS and the $K$-CIFC-CoMS differ thus in the encoding at the cognitive transmitters.
In particular, the $K$-CIFC-CMS consists of encoding functions 
$f_i^{(N)} : [1:2^{N R_1}] \times \ldots \times [1:2^{N R_i}] \to \mathcal{X}_i^N$
such that $X_i^N := f_i^{(N)}(W_1,\ldots,W_i)$, for $i\in[1:K]$, while
all the rest is as for the $K$-CIFC-CoMS.

\subsection{The Gaussian Channel}
\label{sec:ChModel:awgn}
The single-antenna complex-valued $K$-CIFC with Additive White Gaussian noise (AWGN) has input-output relationship
\begin{subequations}
\begin{align}\label{eq:gaussianoutput}
Y_\ell = \sum_{i\in[1:K]} h_{\ell i}X_i+ Z_\ell, \ \ell\in [1:K], 
\end{align}
where, without loss of generality, the inputs are subject to the power constraint
\begin{align}
\mathbb{E}[|X_i|^2] \leq 1, \ i\in [1:K], 
\end{align}
and the noises are marginally proper-complex Gaussian random variables with parameters
\begin{align}
Z_\ell\sim \mathcal{N}(0,1),\  \ell\in [1:K].
\end{align}
\label{eq:AWGN channel model}
\end{subequations}
The channel gains $h_{ij}$, $(i,j)\in[1:K]\times[1:K]$, are assumed constant for the whole codeword duration and therefore known to all terminals.%
This is equivalent to assuming a non-fading / static channel (for which all channel gains can be learnt by every node to any degree of accuracy without impacting the transmission rates as the blocklength tends to infinity~\cite[Chapter 3]{ElGamalKim:InfoTheory}), or to a fading channel with perfect instantaneous channel state information at all nodes~\cite[Chapter 23]{ElGamalKim:InfoTheory}.

Determining the exact capacity region (a convex set in $\mathbb{R}^K_+$) as a function of $K^2$ complex-valued parameters (the channel gains) is a formidable task. Moreover, it is well known that in general different achievable schemes are needed depending on the relative strength of the desired signal at a receiver and the strength of the interfering terms~\cite[Chapter 6]{ElGamalKim:InfoTheory}; determining such regimes is still an art in many cases. To circumvent these problems, in the past decade it has become apparent that {\it it is easier to approximate}~\cite{Tse_approximate}. The approach is as follows.
One first studies a deterministic / noiseless approximation of the Gaussian channel at high SNR, in which the noise is neglected to focus solely on the interference problem. The key is to choose a deterministic model for which capacity can be easily determined but that still retains the distinguishing features of the Gaussian channel. Intuitions from such a well chosen noiseless model can be translated into outer and inner bounds for the Gaussian case at any finite SNR, with the property that the worst case gap /difference between the outer and the inner bounds, taken over all possible channel gains, is a (small hopefully) constant.

\subsection{The Linear Deterministic Approximation (LDA) of the Gaussian Channel at High SNR}
\label{sec:ChModel:LDA}
The Linear Deterministic Approximation (LDA) of Gaussian noise $K$-CIFC  has input-output relationship~\cite{AVHESTIMERPHD}
\begin{align}
Y_\ell = \sum_{i\in[1:K]} \mathbf{S}^{m-n_{\ell i}} X_i,  \ \ell\in [1:K], 
\label{eq:LDA channel}
\end{align}
where $m := \max\{n_{ij}\}$, 
$\mathbf{S}$ is the binary shift matrix of dimension $m$ (made of all zeros except for the first lower diagonal),
all inputs and outputs are binary column vectors of dimension $m$,
the summation is  bit-wise over of the binary field, and 
the channel gains $n_{\ell i}$ for $(\ell,i)\in[1:K]^2,$ are non-negative integers. 
The channel in~\eqref{eq:LDA channel} can be thought of as the high SNR approximation of the channel in~\eqref{eq:AWGN channel model} with their parameters related as $n_{ij} = \lfloor \log(1+|h_{ij}|^2 ) \rfloor, \ (i,j)\in[1:K]^2$~\cite{AVHESTIMERPHD}.
The model in~\eqref{eq:LDA channel} can be `played with' without much network information theory knowledge, by simply reasoning in terms of recovering bits at each receiver from the received linear combinations of the transmitted bits. For the LDA, linear schemes often turn out to be optimal~\cite[Chapter 3]{AVHESTIMERPHD}.

\begin{table*}
\centering
\caption{Capacity Results for Different Cognitive Interference Channel Models.} 
\label{Table1}
\scalebox{0.8}{
\begin{tabular}{| l | l | l | l |}
\hline
Channel Model  & Capacity Results & Remarks  & References  \\ 
\hline
\vtop{\hbox{\strut $K$-CIFC-CMS}\hbox{\strut Gaussian}}& \vtop{\hbox{\strut Characterization of gDoF and sum-capacity of fully}\hbox{\strut symmetric $K$-user Gaussian to with constant gap }}  & \vtop{\hbox{\strut MIMO Broadcast DPC scheme with one encoding order }\hbox{\strut (K-1) PMS + one global cognitive user sufficient to obtain outer bound}}  &\cite{Maamari_JSAC}, Thm.~\ref{thm:outer K CMS}  \\
\hline
\vtop{\hbox{\strut $K$-CIFC-CMS}\hbox{\strut LDA}}&  Sum-capacity of fully symmetric $K$-user & Scheme requires messages equivalent to CoMS channel &\cite{Maamari_JSAC}  \\
\hline
\vtop{\hbox{\strut $K$-CIFC-CMS}\hbox{\strut Gaussian and General Memoryless }} &  \vtop{\hbox{\strut  Sum-capacity general (non-symmetric) channel }\hbox{\strut  under strong channel gain conditions  }} & Joint Decoding Scheme &\cite{Maamaripmsstrong}   \\
\hline
\vtop{\hbox{\strut $K$-CIFC-CoMS}\hbox{\strut Gaussian and General Memoryless }}&  \vtop{\hbox{\strut  Sum-capacity general (non-symmetric) channel  }\hbox{\strut  under strong channel gain conditions  }}   & Joint Decoding Scheme &\cite{Maamaripmsstrong}  \\
\hline 
\vtop{\hbox{\strut $3$-CIFC-CoMS}\hbox{\strut Gaussian and General Memoryless }} &\vtop{\hbox{\strut  Capacity region for a 3 user channel}\hbox{\strut  under strong channel gain conditions  }}   & Compound MAC scheme&\cite{mahtab,Myungstron} \\
\hline
\vtop{\hbox{\strut IFC+CR}\hbox{\strut LDA}}& \vtop{\hbox{\strut Characterization of capacity region}\hbox{\strut (regions in yellow, red and green in Fig.~\ref{fig:shadedregion})}} & \vtop{\hbox{\strut Capacity still open in moderately weak }\hbox{\strut and weak from cognitive relay (region includes the blue region in Fig.~\ref{fig:shadedregion}) }} &\cite{DytsoCognitiverelay} \\
\hline 
\vtop{\hbox{\strut IFC+CR}\hbox{\strut Gaussian}}& Characterization of capacity region &Capacity still open in moderately weak and weak from cognitive relay  &\cite{Rini:CIFC-CR, DytsoCognitiverelay} \\
\hline
\vtop{\hbox{\strut $3$-CIFC-CMS}\hbox{\strut LDA}}& \vtop{\hbox{\strut Characterization of sum-capacity region}\hbox{\strut (regions in yellow, red and green in Fig.~\ref{fig:shadedregion})}}
&  \vtop{\hbox{\strut IFC+CR message knowledge structure is sufficient to obtain}\hbox{\strut CMS outer bound}} &Thm.~\ref{thm:IFC+CR=CoMS}  \\
\hline 
\vtop{\hbox{\strut $3$-CIFC-CMS}\hbox{\strut Gaussian}}& Characterization of sum-capacity under strong conditions&  \vtop{\hbox{\strut IFC+CR message knowledge structure is sufficient to obtain}\hbox{\strut CMS outer bound}} &Thm.~\ref{thm:strongGaussianCMSequalsIFC+CR}  \\
\hline
\vtop{\hbox{\strut $K$-CIFC-CMS}\hbox{\strut Gaussian}}& Characterization of sum-capacity to within constant gap &  \vtop{\hbox{\strut IFC+CR or CoMS message knowledge structure are sufficient}\hbox{\strut to obtain CMS outer bound }} &Thm.~\ref{thm:K}  \\
\hline
\end{tabular}}
\end{table*}

\subsection{Performance Metrics: Sum-Capacity, Generalized Degrees of Freedom, and Constant Gap Approximation}

In this work we shall focus primarily on achieving the sum-capacity, or throughput, for the Gaussian channel model by leveraging results 
for the LDA $K$-CIFC-CMS. 

The throughput is often used as a performance metric of interest from a network operator point of view, where the revenue is assumed to be proportional to the total delivered traffic irrespective of which receiver actually obtains / pays for the bits. This is of course just one performance measure, that unfortunately  neglects important issues such as fairness among users. The sum-capacity should thus simply be thought of here as a `summary' of the capacity region, the latter  which gives us the ultimate complete network performance characterization with all the involved tradeoffs among competing users.

As finding the exact capacity of a Gaussian network is challenging, it is helpful to first study asymptotic approximations of the sum-capacity from the LDA. Towards this goal, it is convenient to introduce
the notion of {\it Generalized Degrees of Freedom} (gDoF) of a symmetric network~\cite{etw}. 
The gDoF 
is meant to capture the
behavior of the capacity for different relative rates of growth of the interference links compared to the direct links, which is typical of the wireless channel~\cite{etw},  when the network is not noise-limited. The symmetric assumption is made so as to reduce the number of parameters in the network model.

Let $\mathsf{SNR}$ and $\mathsf{INR}$ be non-negative numbers, which are intended to characterize the signal and interference to noise ratios, respectively. 
Let us parameterize the magnitude of the $K^2$ channel gains in~\eqref{eq:gaussianoutput} as 
\begin{subequations}
\begin{align}
|h_{i i}|^2    &:= \mathsf{SNR}, \ i\in [1:K], \\
|h_{\ell i}|^2 &:= \mathsf{INR}= \mathsf{SNR}^{\alpha}, \ (\ell,i)\in[1:K]^2, \ell\not= i,
\end{align}
\label{eq:ch parm}
\end{subequations}
for some non-negative real-valued $\alpha$. 
The phases of the channel gains are assumed to be such that any submatrix of the channel matrix $\mathbf{H}=[h_{i,j}]$ is full-rank~\cite{TSERATELIMITEDTXCOOP}.
The gDoF is
\begin{align}
d(\alpha) &:= \lim_{\mathsf{SNR}\to+\infty} \frac{C_{\Sigma}}{\log(1+\mathsf{SNR})},
\label{eq:def gdof}
\end{align}
where $C_{\Sigma}:=\max\{R_1+\ldots+R_K\}$ is the sum-capacity optimized over all achievable rate tuples.

The special case $\alpha=1$ is referred to as Degrees of Freedom (DoF); the DoF provides a high SNR (or interference limited) approximation of the sum-capacity of the network when all channel gains scale at the same rate, or alternatively, when there is an average power constraint $P$ at all nodes, the channel gains are kept fixed and $P$ is increased to infinity.

The gDoF of the Gaussian channel and the sum-capacity of the LDA may be related as follows. Although not proved in general, so far it has been the case (except possibly for $\alpha=1$) that $C_{\Sigma-\text{LDA}}(\alpha)/n_\text{d} = d(\alpha)$ where $C_{\Sigma-\text{LDA}}(\alpha)$ is the sum-capacity of the symmetric LDA with $n_{ii} = n_\text{d}$ and $n_{ij} = \alpha \ n_\text{d}, \ \forall j\not=i$. Intuitively, the gDoF in~\eqref{eq:def gdof} counts how many equivalent independent interference-free streams can be reliably sent across the network simultaneously. For example, in the classical 2-IFC without cognition (see Fig.~\ref{fig:PMSinCMS}(A) and~\cite[Chapter 6]{ElGamalKim:InfoTheory}), $d(\alpha)\in[1/2,1]$, where $d(\alpha)=1/2$ means that each user can send one interference-free stream for half of the time (i.e., time division is optimal at high SNR), while $d(\alpha)=1$ means that each user at each point in time can send one stream as if it was alone on the network~\cite{etw}. 
.

Knowing the gDoF of the channel amounts to correctly establishing the optimal scaling of the sum-capacity at high SNR, that is, for $\mathsf{SNR}\gg1$ we have $C_{\Sigma} \approxeq d(\alpha) \cdot \log(\mathsf{SNR})$; for this reason the gDoF  is also known as ``pre-log factor'' or ``multiplexing gain.'' 
So far it has been the case that the characterization of the gDoF allows one to make an educated guess on {\it good} outer an inner bounds for the original Gaussian channel. 
Specifically, from the study of the LDA one infers how to enhance the original Gaussian channel (by giving the nodes `genie side information' for example) so that the capacity of this enhanced model, $\overline{C}_{\Sigma}(\mathbf{H})$, can be easily determined and thus provides an upper bound to the sum-capacity of the original channel with channel gains $\mathbf{H}=[h_{i,j}]$. 
At the same time, one mimics the LDA capacity achieving scheme and derives a lower bound, $\underline{C}_{\Sigma}(\mathbf{H})$, to the sum-capacity. The goodness of the bounds 

is measured by their {\it additive gap}. Let $\mathsf{gap} := \sup_{\mathbf{H}}(\overline{C}_{\Sigma}(\mathbf{H})-\underline{C}_{\Sigma}(\mathbf{H}))$. If $\mathsf{gap} < +\infty$ we say that the capacity is approximately known to within $\mathsf{gap}$ bits per channel use. Notice that the $\mathsf{gap}$ holds for all channel gain matrices $\mathbf{H}$ and represents the difference between outer an inner bounds for the worst channel in the Gaussian family.

\section{Survey}
\label{sec:survey}

After having introduced the formal channel model definition and the performance metric of interest, we are ready to summarize known results for the CIFC. A summary of the main capacity results mentioned in this survey for Gaussian, LDA and general memoryless channels with the corresponding references are presented for the reader convenience in Table~\ref{Table1}.

\subsection{The Two-User Cognitive Interference Channel ($2$-CIFC)}
\label{sec:survey:2CIC}

The 
overlay two-user cognitive radio channel, otherwise known as the the $2$-CIFC, first introduced in~\cite{devroye_IEEE} (and described in layman terms in magazine articles~\cite{Devroye:commag, Devroye:SPmag}), models the communication in a network between a licensed user (that has the exclusive right to transmit) and a secondary user that has knowledge of the primary user's codebook as well as its message prior to transmission.  
With these idealized `side information' assumptions, the secondary transmitter can act selfishly, pre-cancel the effect of the primary's interference and transmit its own independent message by using dirty paper coding (DPC) in order to cancel the effect caused by the interference~\cite[Section~7.7]{ElGamalKim:InfoTheory}%
\footnote{DPC, or Gelfand-Pinsker binning for channels with states~\cite{gelfand},~\cite[Chapter 3]{ElGamalKim:InfoTheory}, is a coding technique to convey data through a channel subjected to an interference known non-causally to the transmitter but unknown to the receiver. The technique, sometimes called precoding, in the Gaussian noise channel with additive interference is such that the achievable rate is as if the receiver knew the interference as well and were able to perfectly subtract it from the received signal, thus resulting in the same capacity as if the interference was not present~\cite{costa}. DPC can be seen as a building block of Marton's achievable region for the broadcast channel~\cite[Section~8.3]{ElGamalKim:InfoTheory} and is capacity achieving for the $K$-user MIMO Gaussian broadcast channel~\cite{weingarten_MIMOBC}.},
or it can act selflessly and behave as a relay by devoting some of its power to beam-form the primary's message.

The CIFC is a channel model that has elements of {\it both} a broadcast channel~\cite[Chapters 5,8]{ElGamalKim:InfoTheory} and an interference channel~\cite[Chapter 6]{ElGamalKim:InfoTheory}; therefore, in addition to DPC and beam forming, schemes such as rate splitting, superposition coding, and simultaneous non unique-decoding have been devised for the $2$-CIFC \cite{rini:journal1}.\footnote{
Rate splitting~\cite[Section~6.5]{ElGamalKim:InfoTheory} refers to a technique where a user divides its message into independent substreams and possibly codes them with different techniques. 
Superposition coding~\cite[Section~5.3]{ElGamalKim:InfoTheory} is a technique originally devised for the 2-user broadcast channel where a `cloud center' codeword is intended to be decoded by the weakest receiver, while both the `cloud center' codeword and the `satellite' codeword are decoded by the strongest receiver; these techniques allows  users with different channel qualities to decode as much information as they can without becoming the network bottleneck.
Simultaneous non unique-decoding~\cite[Section~6.5]{ElGamalKim:InfoTheory} can be informally thought of 
as giving a receiver the possibility to decode a non-intended / ``don't care'' message if that helps to achieve a larger rate for the intended message~\cite{BandemeretalAllerton2012}. 

The advantage of simultaneous non unique-decoding in a superposition coding scheme with rate splitting is that the 
receivers can decode only part of the interference and thus the resulting achievable rate region is described by fewer rate constraints than more traditional coding techniques, as showed in~\cite{motani}.%
}

Since its introduction in~\cite{devroye_IEEE}, numerous capacity and approximate capacity results have emerged. In particular~\cite{pramodweak} and~\cite{wu:ifcdms} determine the sum-capacity in ``weak interference'' {(i.e., $\mathsf{SNR}>\mathsf{INR}$ in the symmetric Gaussian case)}, while~\cite{maric_uni} characterizes the capacity under ``strong interference'' {(i.e., $\mathsf{SNR}\leq\mathsf{INR}$ in the symmetric Gaussian case)}. The most comprehensive results on the $2$-CIFC can be found in~\cite{rini:journal1,rini:journal2}, where~\cite{rini:journal1} provides the largest achievable rate region for general memoryless channels and some new capacity results, and~\cite{rini:journal2} obtains the capacity to within one bit for the additive white Gaussian noise $2$-CIFC. 
In particular, the key technical aspect for determining the capacity to within a constant gap for the Gaussian $2$-CIFC is to compare the ``unifying outer bound'' in~\cite[Theorem 6]{rini:journal1} to the ``Scheme (C)'' (inspired by Marton's achievable region for the broadcast channel) in~\cite[eq.(21)]{rini:journal2} with the auxiliary random variable assignment in~\cite[eq.(22)]{rini:journal2}, which was inspired by the ``New Capacity Result for the Semi-Deterministic Channel'' (i.e., the signal at the cognitive receiver is an arbitrary deterministic function of the channel inputs) in~\cite[Theorem 11]{rini:journal1}. With this, it is a matter of simple entropy computations to find out that the gap for a user is equal to its number of receive antennas.

The performance of the multi-input and multi-output (MIMO) $2$-CIFC was examined in~\cite{Sriram_MIMO_Cog}, and results were further
refined in~\cite{rini:JSAC2014}, where the capacity of the MIMO $2$-CIFC was derived to within an additive and multiplicative gap proportional to the number of antennas at the secondary receiver. While a finite additive gap (i.e., difference between outer and inner bounds) is meaningful at high SNR, a finite multiplicative gap (i.e., ratio between outer and inner bounds) is meaningful at low SNR--see also the related notion of `wideband slope'~\cite{verduwidebandslope,tuninetti_caire_verdu:isit2002}.

In~\cite{Jafar_MIMO_Cog}, the DoF for the MIMO $2$-IFC with different combinations of non-causal message knowledge at the transmitters and the receivers was obtained.
We note that contributions on the $2$-CIFC extend beyond those mentioned in this survey. The work in~\cite{rini:JSAC2014} provides a more comprehensive overview of the different outer bounds, achievability schemes and capacity results for different channel models of the $2$-CIFC, including the Gaussian noise model.

\subsection{The Three-User Cognitive Interference Channel ($3$-CIFC)}
\label{sec:survey:3CIC}

Several $3$-user extensions of the $2$-CIFC model have been studied in the literature that are the main focus of this paper. For the $3$-user case, which we term the $3$-CIFC, the work~\cite{nagananda2011} proposed the following models for cognition/message sharing: 
\begin{itemize}
\item  with {\it cumulative message sharing} ($3$-CIFC-CMS)
\item  with {\it primary message sharing} ($3$-CIFC-PMS)
\item  with {\it cognitive-only message sharing} ($3$-CIFC-CoMS). 
\end{itemize}

The $3$-CIFC-CMS, shown 
in Fig.~\ref{fig:PMSinCMS}(C), models a network of two cognitive users: Tx$_2$ knows the message of primary user Tx$_1$, and Tx$_3$ knows the messages of Tx$_1$ and Tx$_2$. The $3$-CIFC-PMS is shown in Fig.~\ref{fig:PMSinCMS}(B), in this network there are 2 cognitive users: Tx$_2$ and Tx$_3$, who only know the message of the primary Tx$_1$. 
In the $3$-CIFC-CoMS, shown in Fig.~\ref{fig:PMSinCMS}(E), there are two primary users who do not know each others' message and a single cognitive user who knows both primary messages.  In Fig.~\ref{fig:PMSinCMS}(D) we have also plotted an Interference Channel with a Cognitive Relay (IFC+CR), a channel model studied in~\cite{Rini:CIFC-CR, cognitiverelaysriram, DytsoCognitiverelay}, in which there are two primary Tx$_1$ and Tx$_2$ and one cognitive Tx$_3$ which has knowledge of the two primary messages and aids in their transmission. This third node is a relay only, as it does not have a message of its own to transmit. 
Clearly, because the channel with CMS has more message knowledge at its transmitters, anything the channel with PMS can do, the channel with CMS can as well. Hence, in Fig.~\ref{fig:PMSinCMS}, we have used the $\subseteq$ to denote that the capacity region of the PMS is contained in that with the CMS message knowledge structure.

Limited prior work has emerged on the $3$-CIFC: in~\cite{nagananda2009information}, the authors considered the $3$-CIFC-PMS and CMS, achievable rate regions based on rate splitting and binning were derived. The $3$-CIFC-CoMS scenario was later introduced in~\cite{nagananda2010achievable}. Achievable rate regions for the discrete memoryless channel were obtained and were numerically evaluated for the Gaussian noise channel; a numerical comparison of the achievable rates in Gaussian noise was made in~\cite{nagananda2011}.
Inner and outer bounds for the special case of the $3$-CIFC-CoMS in which the cognitive user is assumed not to interfere with the primary users were obtained in~\cite{mirmohseni2011capacity}. 
he work in~\cite{mahtab,Myungstron} also considered the $3$-CIFC-CoMS and provided the capacity region under strong interference, where the channel reduces to a compound Multiple Access Channel (MAC).\footnote{A MAC models the uplink of a cellular system with multiple transmitters with mutually independent messages and a single receiver interested in decoding all messages~\cite[Chapter~4]{ElGamalKim:InfoTheory}. A compound MAC has multiple receivers, all interested in decoding all messages. The compound channel is often used to model channel state information uncertainty at the transmitters.}
The results were evaluated for the Gaussian noise channel. 

We note that an achievable region for the $3$-CIFC-CoMS can be obtained by considering the IFC+CR, where a `selfless' cognitive node does not send any data to its receiver and instead manages the interference in the network. The most comprehensive results on the symmetric LDA of the IFC+CR can be found in~\cite{DytsoCognitiverelay}. We shall see later on that there exist channel gain conditions for which the sum-capacities of the IFC+CR and of the $3$-CIFC-CoMS are  to within a constant gap, which implies that sometimes the most throughput-wise efficient use of cognitive abilities is to manage interference.

However, general relationships between the capacity regions of the three channel models has not yet been examined. In this work we make progress by showing some relationships between the capacity regions of the different models. This has implications in understanding which message knowledge structures are most desirable in practice. 

For example, it turns out that having message knowledge at the transmitters--corresponding to a PMS or a CoMS--is sufficient for achieving (under certain channel gain conditions, and to within constant gap) the sum-capacity upper bound derived for the CIFC-CMS. This has obvious practical advantages and may reduce the amount of signaling needed in practice on the backhaul in order to achieve the desired message knowledge structure.

\subsection{The $K$-user Cognitive Interference Channel ($K$-CIFC)}
Much less work has been done on cognitive interference channels with an arbitrary ($K$) number of transmit and receive pairs. Given the large number of channel gain parameters involved, and the many ways messages may be shared at the transmitters, studies have so far usually been restricted to symmetric scenarios, or have even assumed some links to be zero altogether. 
For example, in~\cite{vishwanathjafarianmultiusercognitive} the authors considered a channel model that consists of one primary user and $K-1$ cognitive users, a $K$-user extension of the PMS scenario: each cognitive user only knows the primary message in addition to their own message. However, they restrict the channel so that the cognitive users do not cause interference to one another but only to the primary receiver and are interfered only by the primary transmitter; for this channel model the capacity in the very strong interference regime {(i.e., $\mathsf{INR} > \mathsf{SNR}^2$ in the symmetric case)} is obtained by using lattice codes.%
\footnote{An $n$-dimensional lattice is a discrete additive group in $\mathbb{R}^n$. The lattice points $x$ that satisfy the power constraints $\|x\|^2 \leq nP$ can be used as a code for the point-to-point Gaussian noise channel. Shannon's capacity theorem says that an optimal block code for a bandwidth-limited Gaussian channel consists of a dense packing of code points within a sphere in a high-dimensional Euclidean space~\cite{Shannon:1948}; since most of the densest known packings are lattices, it makes sense to study the capacity achieving properties of lattices-based codes. Lattice codes constructions are known to approach $1/2\log(1+\mathsf{SNR})$~\cite{Erez:2004:latticecapacity}, and in general lattice codes are {\it good for almost anything}~\cite{erez2005lattices}. Lattice codes are also a key ingredient for the so-called compute-and-forward protocol in relay networks~\cite{nazer2009compute}.

It was shown that, under certain channel gain conditions, users could simultaneously communicate as if the network experiences no interference. In~\cite{Maamaripmsstrong} the sum-capacity of the fully connected $K$-CIFC-CoMS and a $K$-CIFC-CMS under certain strong interference conditions was derived; it was shown that simply beam-forming to the primary receiver was sum-capacity optimal.}

In~\cite{Maamari_JSAC}, we further considered the $K$-CIFC-CMS. A sum-capacity upper bound was derived by giving {\it nested} genie side-information to receivers. The side information given to the secondary receivers consists of messages and output sequences of receivers with lesser index. It was shown that this upper bound can be achieved exactly for the symmetric linear deterministic $K$-CIFC. 

For the Gaussian $K$-CIFC the capacity was derived to within an additive and multiplicative gap independent of the channel gains. Interestingly, for the symmetric $K$-CIFC-CMS, the achievability scheme required only cognitive message knowledge at one user, a global cognitive user. In particular, the message knowledge sufficient to achieve the upper bound corresponds to that of the $K$-CIFC-CoMS.  The sum-capacity for the Gaussian $K$-CIFC-CMS was then characterized, in which the proposed achievability scheme was inspired by the dirty paper coding region of the MIMO broadcast with one encoding order. After carefully choosing the power splits so as to match the sum-capacity upper bound, it turned out that the message knowledge sufficient to achieve the sum-capacity (to within a constant gap from the derived upper bound) can be simplified to that of a $(K-1)$-CIFC-PMS and only one global cognitive user, whose role is to simultaneously ``manage'' all the interference in the network. 

For the remainder of the paper, we will be interested in investigating the benefit of a cumulative message sharing structure (the capacity of a $K$-CIFC-CMS) over that of a cognitive channel with a cognitive-only message sharing structure (the capacity of the $K$-CIFC-CoMS), which requires much less message knowledge. It turns out that in some cases, the difference in sum-capacity between the two types of message knowledge structure is only a constant gap. The remainder of the paper elaborates on this by first 
providing past work on the degrees of freedom results for such networks.

\subsection{Past Work on gDoF for Multi-user Cognitive Networks}

Recall that the DoF (or gDoF for {$\alpha=1$}) of a network provides a high SNR / interference-limited approximation of the sum-capacity of that network, and is usually easier to characterize than the capacity at finite SNR.  

As a baseline, the DoF of the $K$-IFC without any cognitive user is almost surely $\text{DoF}=K/2$~\cite{Jafar:2008:alignment,motahari2009real} if the channel gains are {\it generic}, that is, drawn independently at random from a continuous distribution without a mass at zero (the Rayleigh distribution for example).
Moreover, when source and destination nodes are distinct, {\it strictly causal} feedback, node cooperation, and relaying cannot increase the DoF above that of the classical $K$-IFC~\cite{Jafar:2009:relays}, which motivates the study of non-causal / cognitive networks. 
For the $2$-user case, the DoF with cognition is given in~\cite[Eq.1]{Jafar:cognition}.
We note also that the DoF of (non-generic) networks with a sparse topology, i.e., networks for which some channel gains are zero, cannot be smaller than that of the classical $K$-IFC.

In~\cite{compvenu}, the authors characterized the DoF of a $K$-IFC in which each transmitter, in addition to its own message, has access to a subset of the other users' messages; in particular, it was shown that the maximum possible $\text{DoF}=K$ is attainable if the sum of the number of jointly cooperating transmitters and the number of jointly decoding receivers is greater than or equal to $K+1$. 
In general, cognition 
cannot decrease the DoF over that of the classical $K$-IFC. The question of interest is thus how much cognition is needed to `beat' the classical $K$-IFC, or alternatively, for a given DoF above that of the classical $K$-IFC what is the minimum amount of cognition needed to achieve it. The novel results in this paper try to answer these questions.

The DoF ($\alpha=1$) of the  
$K$-IFC in which receiver $k$ suffers from interference due to the transmission of users indexed $i\leq k$ only, and where transmitter $k$ knows the messages of $J$ preceding transmitters, was characterized in~\cite{Wigger_cog} as\begin{align*} 
{ \rm DoF}(K,J) = 1-\frac{1}{K}\left \lfloor{\frac{K}{J+2}}\right\rfloor .
\end{align*} 
The achievability scheme utilizes DPC at the cognitive transmitters indexed $i\in[1:J+1]$ while silencing transmitter $J+2$ (repeated at all users).

\begin{figure}
\centering
\includegraphics[width=0.47\textwidth]{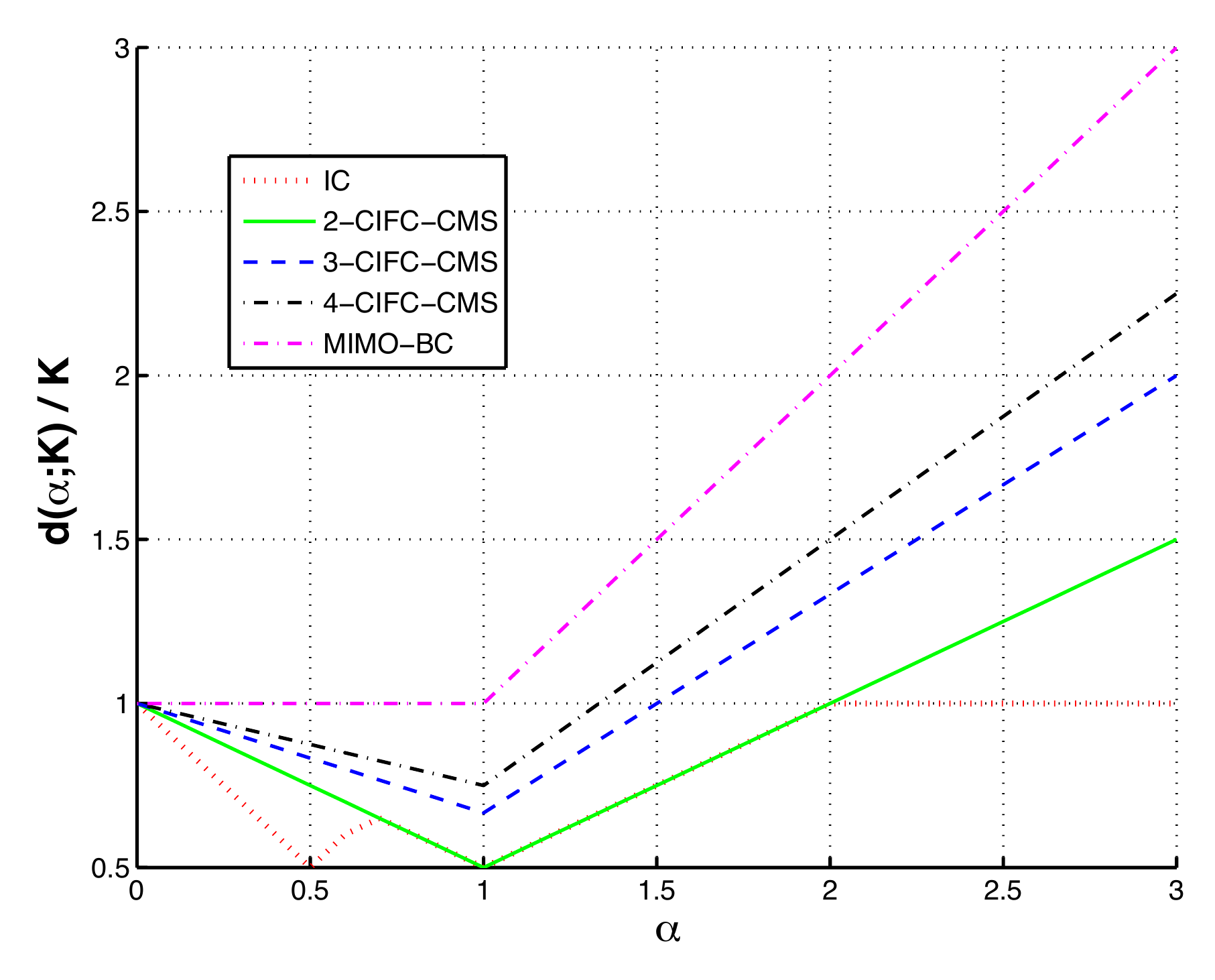}%
\caption{Normalized Generalized Degrees of Freedom of the $K$-user MISO broadcast, interference, CIFC-CMS channels (normalized by the number of transmitters).} 
\label{fig:GDof}%
\end{figure}

In~\cite{Maamari_JSAC}, the gDoF of the $K$-CIFC-CMS was shown to be
\begin{align*}
d^{\text{K-CIFC-CMS}}(\alpha)= K \max\{1,\alpha\}-\alpha.
\end{align*} 
In Fig.~\ref{fig:GDof} the gDoF of the $K$-CIFC-CMS for $K\in[2:4]$ is plotted along with the gDoF of the MISO broadcast channel, which provides an upper bound by giving cognitive message knowledge of all messages at all transmitters, given by~\cite{Jafar_DOF_IC} .
\begin{align*}
d^{\text{K-BC}}(\alpha) = K \max\{1,\alpha\},
\end{align*}
and the gDoF of the $K$-IFC, 
which provides a lower bound 
where transmitters have no cognitive abilities, given by~\cite{Jafar_DOF_IC}
\begin{align*}
d^{\text{K-IFC}}(\alpha) = K \min\left\{
1,
\max\left(\frac{\alpha}{2},1-\frac{\alpha}{2}\right),
\max\left(\alpha,1-\alpha\right)
\right\}.
\end{align*}

Two interesting observations can be made on the normalized (by the number of users) gDoF~\cite{Maamari_JSAC}:
the first is that unlike the $K$-IFC and the MISO broadcast channels, the normalized gDoF of the $K$-CIFC-CMS is a function of the number of users in the networks $K$, and 
the second is that the normalized gDoF loses $\alpha/K$ with respect to the $d^{\text{K-BC}}(\alpha)/K$ (a vanishing loss as $K$ increases).

\medskip
We have now concluded the survey on known results on multi-user $K$-CIFC. We are thus ready to present our novel results. We start by presenting an outer bound in Section~\ref{sec:NEWouter} for the general memoryless channel (defined in Section~\ref{sec:ChModel:general:CIFC-CoMS and CMS}), which we then specialize in Section~\ref{sec:NEWlda} to the LDA (defined in Section~\ref{sec:ChModel:LDA}) and finally in Section~\ref{sec:NEWawgn} to the Gaussian channel (defined in Section~\ref{sec:ChModel:awgn}).

\section{Novel Outer Bounds}
\label{sec:NEWouter}

We start by reporting an outer bound region for the general memoryless $K$-CIFC-CoMS as defined in Section~\ref{sec:ChModel:general:CIFC-CoMS and CMS}. 
Based on the $K$-CIFC-CoMS result, we then derive an outer bound for the $3$-CIFC-CMS.
Note that an outer bound on the general memoryless $K$-CIFC-CMS is also an outer bound for the $K$-CIFC-CoMS as the CMS has more message knowledge at the transmitters. One of our goals is to show that for {\it distributed cognition} ($K$-CIFC-CMS) is not always needed from a sum-capacity perspective. In particular, the sum-capacity upper bound derived for the $K$-CIFC-CMS may sometimes be achieved (or achieved to within a constant gap) with a message structure corresponding to that of the $K$-CIFC-CoMS, indicating that extra message knowledge sometimes only leads to bounded gains.

\begin{thm}[$K$-CIFC-CMS Outer Bound~{\cite[Theorem~4]{Maamari_JSAC}}]
\label{thm:outer K CMS}
The capacity region of the general memoryless $K$-CIFC-CMS is contained in the region 
\begin{subequations}\label{eq:outer bound general K progressive}
\begin{align}
   R_i &\leq I(Y_i; X_{i},X_K|X_{[1:i-1]}), 
\\ \sum_{{ j=i}}^{K} R_j &\leq \sum_{j=i}^{K} I(Y_j; X_{[j:K]}| X_{[1:j-1]},Y_{ [1:{j-1}]}),
\end{align}
\label{eq:K outer CoMS}
\end{subequations}
for $i\in[1:K]$
for some joint input distribution $P_{X_1,\ldots,X_K}$.
Moreover, each rate bound in~\eqref{eq:outer bound general K progressive} may be tightened with respect to the joint channel conditional distribution as long as the marginal channel conditional distributions are preserved~\cite{Maamari_JSAC}.
\end{thm}

Since we will be examining the $3$-CIFC-CoMS at length, we explicitly provide the statement and proof for the case $K=3$ below, which is based on Theorem~\ref{thm:outer K CMS} for $K=3$.

\begin{thm}[$3$-CIFC-CoMS Outer Bound]
\label{thm:outer K=3 CoMS}
The capacity region of the general memoryless $3$-CIFC-CoMS is contained in the region defined by
\begin{subequations}
\begin{align}
          R_1 &\leq I(Y_1;     X_1,X_3| X_2),     \label{eq:CoMS 1}
\\        R_2 &\leq I(Y_2;     X_2,X_3| X_1),     \label{eq:CoMS 2}
\\        R_3 &\leq I(Y_3;         X_3| X_1,X_2), \label{eq:CoMS 3}
\\    R_1+R_3 &\leq I(Y_1;     X_1,X_3| X_2)
                  + I(Y_3;         X_3| X_2,X_1,Y_1), \label{eq:CoMS 1,3}
\\    R_2+R_3 &\leq I(Y_2;     X_2,X_3| X_1)
                  + I(Y_3;         X_3| X_1,X_2,Y_2), \label{eq:CoMS 2,3}
\\R_1+R_2+R_3 &\leq I(Y_1; X_1,X_2,X_3) \nonumber
                  + I(Y_2;     X_2,X_3| X_1,Y_1)        \\&
                  + I(Y_3;         X_3| X_1,Y_1, X_2,Y_2), \label{eq:CoMS 1,2,3 a}
\\R_2+R_1+R_3 &\leq I(Y_2; X_2,X_1,X_3) \nonumber
                  + I(Y_1;     X_1,X_3| X_2,Y_2)  \\&       
                  + I(Y_3;         X_3| X_2,Y_2, X_1,Y_1), \label{eq:CoMS 1,2,3 b},
\end{align}
\label{eq:CoMS outer region}
\end{subequations}
for some joint input distribution $P_{X_1,X_2,X_3}$. 
\end{thm}
\begin{IEEEproof}
An outer bound  for the $3$-CIFC-CoMS, where transmitter $i\in[1:2]$ only knows its own message can be obtained by giving side information to the two primary users so as to transform the CoMS message structure in Fig.~\ref{fig:PMSinCMS}(E) into the CMS one in Fig.~\ref{fig:PMSinCMS}(F). For each possible permutation of the primary users' indices we obtain a region as in~\eqref{eq:K outer CoMS}; by intersecting these 
regions we obtain the outer bound for the $3$-CIFC-CMS in~\eqref{eq:CoMS outer region}. 
\end{IEEEproof}

Since we are mainly interested in the sum-capacity, we explicitly derive a sum-capacity upper bound for the $3$-CIFC-CoMS from Theorem~\ref{thm:outer K=3 CoMS} as follows.

\begin{cor}
\label{thm:sum-capacity outer K=3}
The sum-capacity of the general memoryless $3$-CIFC-CoMS is upper bounded by
\begin{subequations}
\begin{align}\nonumber
 & R_1+R_2+R_3 \leq \min(a,b), 
\nonumber
\\& a:= \min\Big\{
                    I(Y_3;         X_3| X_1,X_2,Y_1),
                    I(Y_3;         X_3| X_1,X_2,Y_2)
                  \Big\}
\nonumber 
\\&\quad
+ I(Y_1;     X_1,X_3| X_2)
+ I(Y_2;     X_2,X_3| X_1)
, 
\label{eq:CoMS sumrate outerbound 1}
\\& b:=
\min\Big\{
   I(Y_1; X_1,X_2,X_3)          + I(Y_2;     X_2,X_3| X_1,Y_1), 
\nonumber\\&\qquad\qquad
   I(Y_1;     X_1,X_3| X_2,Y_2) + I(Y_2; X_1,X_2,X_3) 
\Big\}
\nonumber 
\\&\quad
+I(Y_3; X_3| X_2,Y_2, X_1,Y_1),
\label{eq:CoMS sumrate outerbound 2}
\end{align}
\label{eq:CoMS outer sumrate}
\end{subequations}
for some input distribution $P_{X_2,X_1,X_3}$. 
\end{cor}
\begin{IEEEproof}
The sum-rate upper bound in~\eqref{eq:CoMS outer sumrate} is obtained as $\min\{\eqref{eq:CoMS 1,3}+\eqref{eq:CoMS 2}, \eqref{eq:CoMS 2,3}+\eqref{eq:CoMS 1}, \eqref{eq:CoMS 1,2,3 a}, \eqref{eq:CoMS 1,2,3 b}\}$. 
\end{IEEEproof}

We conclude this section by highlighting some of the `desirable characteristics' one seeks in an outer bound that can be found in Theorems~\ref{thm:outer K CMS},~\ref{thm:outer K=3 CoMS}, and Corollary ~\ref{thm:sum-capacity outer K=3}.
Our outer bounds apply to any {\it memoryless} channel (because of no assumptions on the channel structure, as opposed to structure specific bounds such as for `injective semi-deterministic' channels~\cite[Section~6.7]{ElGamalKim:InfoTheory}). As such, they can be used as building blocks for other networks (through a cooperation or a genie side information argument as we did with with Theorem~\ref{thm:outer K CMS} to obtain Theorem~\ref{thm:outer K=3 CoMS}). Last but not the least, they are easily computable (because they do not involve random variables that are not part of the problem definition, but only channel inputs and outputs). For example, our bounds as exhausted by independent and equally likely input bits for the LDA, and by jointly Gaussian inputs for the Gaussian IFC.

In the following we shall derive capacity results that will however not cover all possible parameter regimes or the whole capacity region. We offer next our thoughts as to why this may be the case. 
The LDA and the Gaussian noise channels are examples of Injective Semi-Deterministic (ISD) channels.
ISD~IFCs are such that the channel output of each user is a deterministic function of the intended input and a noisy function of the interferers with the property that, knowing the channel output and the intended input (which is the case after decoding) it is possible to recover the noisy function of the interferers~\cite{telatar_tse}. The ISD characteristics can be leveraged to obtain outer bounds that may not be derived for the general memoryless case. For example, in~\cite{telatar_tse} it was shown that for the ISD 2-IFC the Han and Kobayashi achievable region~\cite[Section~6.5]{ElGamalKim:InfoTheory} is optimal to within a constant gap by deriving ISD-specific sum-rate upper bounds and bounds of the type $2R_1+R_2$ and $R_1+2R_2$. Interestingly, it turns out that such ISD-specific bounds are not needed in order to characterize to within a constant gap the capacity of the $2$-CIFC~\cite{rini:journal2} or that of the $2$-IFC with output feedback~\cite{suh2010feedback}; but they are needed for the IFC+CR~\cite{DytsoCognitiverelay} and for the $2$-IFC with causal cognition~\cite{arXiv:1503.07372}. At this point it is not clear whether our partial novel capacity results for the LDA and the Gaussian noise channels are due to a weakness in the achievable scheme or a lack of ISD-specific bounds, or both.  Answering this question is a subject of current investigation.

\section{Novel Capacity Results for the LDA}
\label{sec:NEWlda}

In this section we shift our focus to the $3$-user LDA, before returning to arbitrary $K$-user Gaussian CIFCs.
As mentioned before, the LDA models the Gaussian channel at high SNR and provides insights into interference-limited behaviors of the network, i.e., how the capacity is limited by the interference created and caused by other users rather than by noise.
We next study both the symmetric LDA in Section~\ref{sec:NEWldaSym} and the non-symmetric LDA in Section~\ref{sec:NEWldaNONSym} with three users. We show that conclusions for symmetric channels may or may not extend to more general asymmetric settings.

\subsection{Sum-Capacity of the Symmetric LDA $3$-CIFC-CMS}
\label{sec:NEWldaSym}

In our prior work, we showed that the sum-capacity of the $3$-CIFC-CoMS is the same as that for the $3$-CIFC-CMS for a symmetric (all cross-over links are the same strength, all direct links the same strength) LDA~\cite[Theorem~4]{Maamari_JSAC}. We now strengthen and generalize this result by obtaining channel conditions under which the sum-capacity of the $3$-CIFC-CMS is actually achieved by a scheme which only requires the message knowledge of an IFC+CR~\cite{Rini:CIFC-CR, cognitiverelaysriram, DytsoCognitiverelay}, that is, setting $R_3=0$ and ignoring the knowledge of message $W_1$ at Tx$_2$ is sum-capacity optimal.

\begin{figure}
\centering
\includegraphics[width=0.47\textwidth]{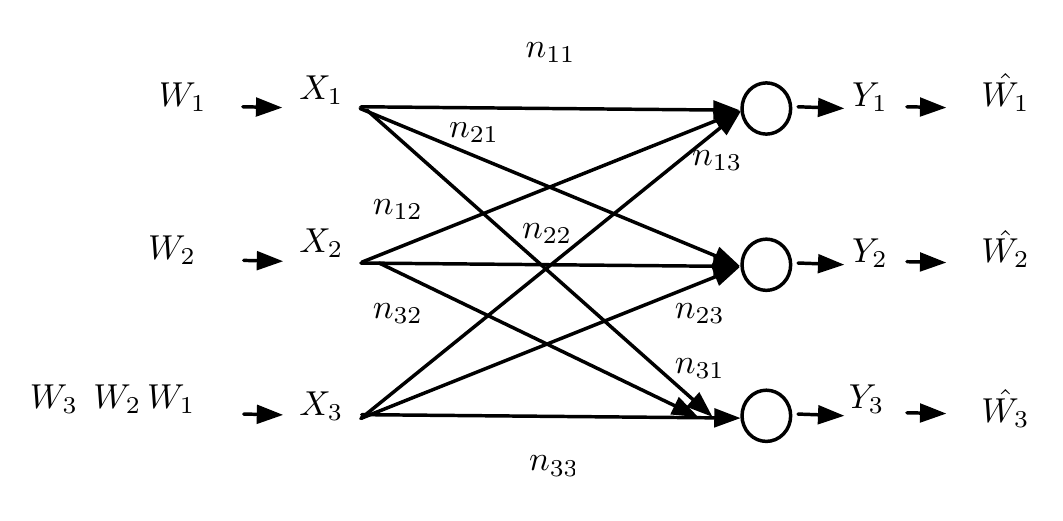}%
\caption{The Linear Deterministic Channel $3$-CIFC-CoMS with channel gains from Tx$_j$ to Rx$_i$ denoted by $n_{ij}$ for $i,j \in [1:3]$. }
\label{fig:LDAchannelgains}
\end{figure}

We show results for the following somewhat symmetric parametrization of the channel gains for the LDA with channel gains as described in~\eqref{eq:LDA channel} and as depicted in Fig.~\ref{fig:LDAchannelgains}
\begin{subequations}
\begin{align}
&n_{22} = n_{11}= n_{\rm d}\\
&n_{12} = n_{21}= n_{\rm i} = \alpha n_{\rm d}\\
&n_{13} = n_{23}= n_{\rm  c} = \beta n_{\rm d}. 
\end{align}
\label{eq:parameterization}
\end{subequations}
Notice that we have not placed any conditions yet on the links coming into the cognitive receiver, i.e., on $n_{31}, n_{32}, n_{33}$. 

Our main result is as follows.
\begin{thm}
\label{thm:IFC+CR=CoMS}
The message knowledge structure of the 
IFC+CR is sufficient to achieve the sum-capacity of the LDA $3$-CIFC-CoMS and LDA $3$-CIFC-CMS channels
under the following channel gain conditions
\begin{align}
& \{ n_{33} \leq \beta n_d\} \; \cup \; \{\alpha \geq 1\}  \; \cup  \; \{\beta \geq \alpha\}. \label{eq:conditionsthm5}
\end{align} 
In addition, the outer bounds for the IFC+CR and those for the $3$-CIFC-CMS (and hence $3$-CIFC-CoMS) coincide for the following channel gains conditions, though it is not generally known whether these outer bounds are achievable
\begin{align}
& \{ n_{33} \leq \beta n_d\} \; \cup \; \{\beta < \alpha < 1\}  \; \cap  \; \{2\leq 3\alpha +\beta\}. \label{eq:conditionsthm5outer}
\end{align}

\end{thm}
\begin{IEEEproof}

The upper bound on the sum-capacity for the LDA $3$-CIFC-CMS in Theorem~\ref{thm:outer K CMS} was evaluated in~\cite[eq.(8)]{Maamari_JSAC} and is given by
\begin{align}\nonumber
R_1+R_2+R_3&\leq \max\{n_{11}, n_{12},n_{13}\} + f(n_{22}, n_{23} |n_{12}, n_{13}) \\&+ [n_{33} - \max\{n_{13},n_{23}\}]^+,\label{eq:sumrateCIFCnonsymmetric}
\end{align}
where the function $f(c,d|a,b)$ in~\eqref{eq:sumrateCIFCnonsymmetric} is defined as $ \max\{c+b,a+d\}-\max\{a,b\} \ \text{if $c-d\not=a-b$}$ and $ \max\{a,b,c,d\}-\max\{a,b\} \ \text{if $c-d=a-b$}$.
In the following we aim to find channel gain conditions under which it is sum-rate optimal for the global cognitive transmitter in a $3$-CIFC-CMS to behave as a cognitive relay. If 

\begin{align}
\label{eq:condition to R3=0}
n_{33}\leq \max\{n_{13},n_{23}\},
\end{align} 
(the condition in~\eqref{eq:condition to R3=0} means that all the $n_{33}$ bits received at Rx$_3$ from Tx$_3$ are also received at Rx$_1$ and Rx$_2$)

and by defining the normalized (by the direct link) rates $r_1 = \frac{R_1}{n_d}$ and $r_2 = \frac{R_2}{n_d}$,
the sum-capacity outer bound in~\eqref{eq:sumrateCIFCnonsymmetric} 

can be re-written as  
\begin{align}\label{eq:sumrateCIFC}
r_1+r_2\leq \max\{\alpha,\beta,1\} + f(1, \beta |\alpha, \beta),
\end{align}
or more explicitly as
\begin{subequations}\label{eq:sumrateCIFCsimplfiied} 
\begin{align}\nonumber
  r_{\rm sum,CIFC-CMS,a}&=r_1+r_2\leq\label{eq:sumrateequation1}
\max\{\alpha,\beta,1\}+ \max\{1,\alpha\}+\\& \beta-\max\{\alpha,\beta\}, \ \text{if $\alpha\not=1$},\\
 r_{\rm sum,CIFC-CMS,b}&=r_1+r_2\leq\max\{\beta,1\}, \ \text{if $\alpha=1$}.\label{eq:sumratealphaequal1}
\end{align}
\end{subequations}
A sum-capacity outer bound for IFC+CR with the same parameterization as in~\eqref{eq:parameterization}  was derived in~\cite[eq.(11)]{DytsoCognitiverelay} as
\begin{subequations}  
\label{eq:sumrateIFC+CR}
\begin{align}
\nonumber
r_{\rm sum,IFC+CR,a}=r_1+r_2&\leq [1-\max\{\beta,\alpha\}]^+\nonumber\\& + \beta +\max\{1,\alpha\},\label{eq:firstsumrateIFC+CR}
\\
r_{\rm sum,IFC+CR,b}=r_1+r_2 &\leq 2\max\{1-\alpha,\alpha,\beta\}\nonumber\\& +2\min\{\alpha,\beta\},\label{eq:secondsumrateIFC+CR}
\\
r_{\rm sum,IFC+CR,c}=r_1+r_2 &\leq \max\{1,\beta\},\ \text{if $\alpha=1$}. 
\label{eq:sumratealpha1}
\end{align}
\end{subequations}
In Appendix~\ref{app:proof of thm IFC+CR=CoMS} we provide the remaining details on how to obtain the channel conditions \eqref{eq:conditionsthm5} and \eqref{eq:conditionsthm5outer} under which \eqref{eq:sumrateequation1} is less than or equal to both \eqref{eq:firstsumrateIFC+CR} and \eqref{eq:secondsumrateIFC+CR}. 
\end{IEEEproof}

\begin{figure}
\centering
\includegraphics[width=0.35\textwidth]{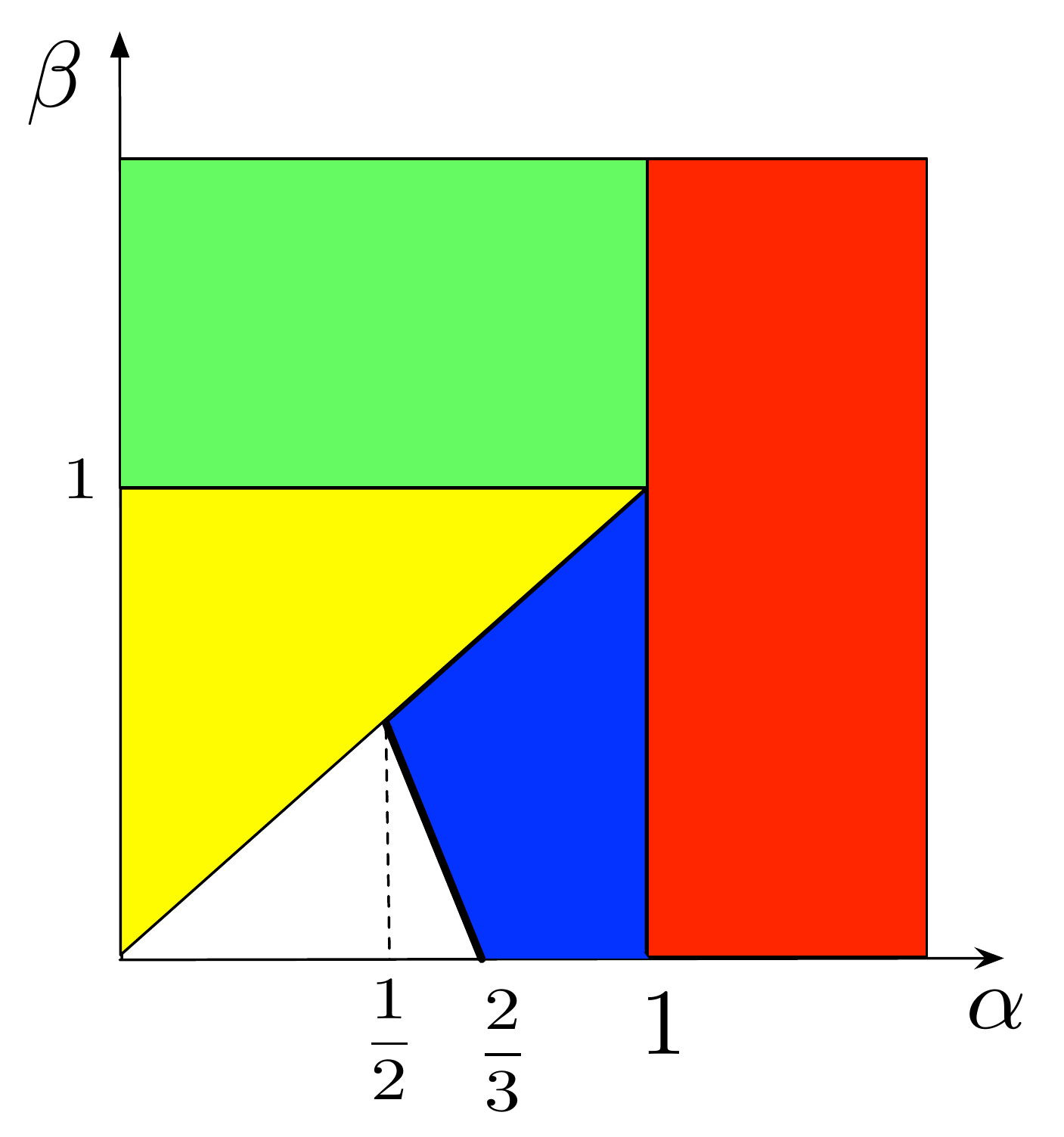}%
\caption{A plot of the channel gain regions defined in~\eqref{eq:conditionsthm5} and~\eqref{eq:conditionsthm5outer} (note that $\{ n_{33} \leq \beta n_d\}$ is not shown). For all regimes (the green and red should extend indefinitely to the top and right, respectively) except the white and blue regions the IFC+CR inner bound achieves the $3$-CIFC-CMS outer bound. In the white regime the known outer bounds do not coincide. In the blue region the outer bounds coincide but it is not known whether these are achievable for the IFC+CR. Therefore, 
for the  white and blue regions it is provably sum-capacity optimal for a cognitive transmitter in the $3$-CIFC-CMS to behave like a cognitive relay.}
\label{fig:shadedregion}
\end{figure}

Let us now discuss what the meaning and implications of Theorem~\ref{thm:IFC+CR=CoMS} are.
A plot of the channel gain relationships obtained in~\eqref{eq:conditionsthm5} and in~\eqref{eq:conditionsthm5outer} is shown in Fig.~\ref{fig:shadedregion}. It is interesting to note that for a wide range of parameter regimes (all but the white and blue regimes) it is sum-capacity optimal for the most cognitive user to simply act as a cognitive relay, by ignoring its own message ($R_3=0$) and only managing interference, and for the 2nd~user to ignore its knowledge of the 1st~user's message. An interesting open question is whether the same holds for the whole capacity region. Regarding what may be missing in the the white and blue regimes to get a sum-capacity result, please recall our discussion at the end of Section~\ref{sec:NEWouter} about ISD-type outer bounds.

Finally, we note that having $R_3>0$ does not qualitatively alter the capacity results. In fact the condition $n_{33} > \max\{n_{13},n_{23}\}$ (i.e., the complement of the condition in~\eqref{eq:condition to R3=0}) suggests the following communication strategy. 
Recall that $n_{33} > \max\{n_{13},n_{23}\}$ means that some of the $n_{33}$ bits received at Rx$_3$ from Tx$_3$ are neither received at Rx$_1$ nor at Rx$_2$, i.e., they do not create interference (commonly referred to as ``bits are received below the noise floor of the non-intended receivers'').
Therefore, the most cognitive user can use its bits not received at Rx$_1$ and at Rx$_2$ to convey its own message to Rx$_3$ while keep using an IFC+CR type scheme to manage interference for  those bits that are received at Rx$_1$ and at Rx$_2$.

We now briefly consider the role of symmetry in our LDA results, and consider examples of $3$-CIFC-CMS with asymmetric channel gains. We ask whether our 
statements on message knowledge and cognitive relay behavior continue to hold then. It turns out that under certain asymmetric channel gains, the statements  continue to hold. In other cases, however, having cognitive transmitters as in a CIFC-CMS is critical to achieve the outer bound in~\eqref{eq:sumrateCIFCnonsymmetric}. We provide examples that illustrate both observations. The characterization of {\it exactly when}---in terms of channel gain relationship---either of the former mentioned schemes is optimal is still an open problem.

\subsection{Sum-Capacity for some Asymmetric LDA $3$-CIFC-CMS}
\label{sec:NEWldaNONSym}

In order to explore whether the conclusions made regarding cognition and message knowledge are due to our restriction to symmetric channel gain conditions,  
we present some examples which achieve the sum-capacity outer bound in~\eqref{eq:sumrateCIFCnonsymmetric} (valid for all channel gains) for the LDA asymmetric $3$-CIFC-CMS. We note that attempting to characterize the sum-capacity for all asymmetric channel gain conditions is an open problem; the number of channel gains that needs to be considered is large and many different relative orderings of these channel gains may need to be considered in general.  

A note on reading the following figures. In the following examples, bits within each transmit signal vector are represented by different shades of the same color. Bits that arrive at the same level (equal shifts) from the different users will {\it neutralize} by addition modulo two over the binary field if they are of the same color, while bits arriving at the same level but with different color interfere. Having knowledge of primary users messages, cognitive users can transmit bits with different colors (corresponding to primary users) and its own bits.

\begin{figure}
\centering
\includegraphics[width=0.47\textwidth]{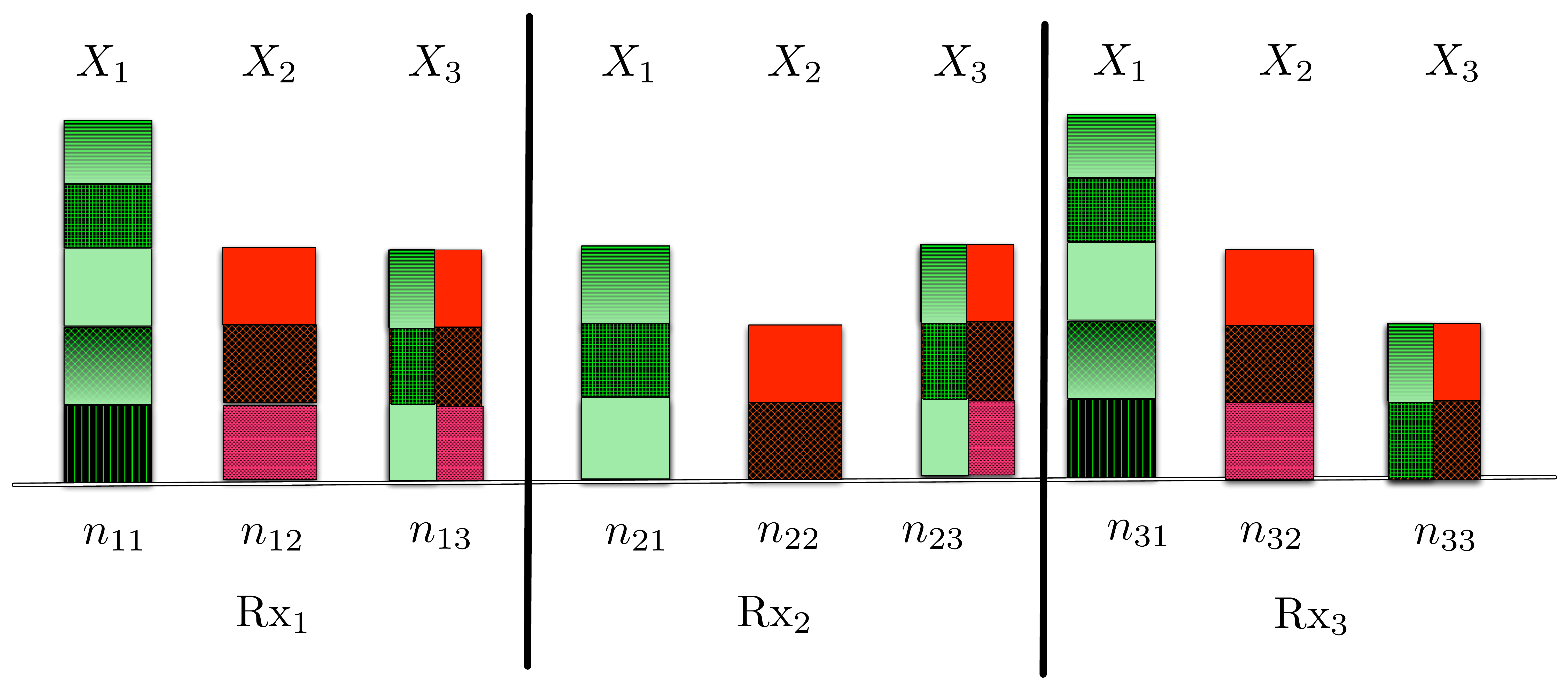}%
\caption{LDA in Example 1. The cognitive transmitter behaves as a cognitive relay zero forcing interference at the primary receivers simultaneously.}
\label{fig:LDA3}%
\end{figure}

\paragraph{Example 1, Fig.~\ref{fig:LDA3}}\label{ex1}
Parameters: $n_{11}=5, n_{12}=3, n_{13}=3, n_{21}=3, n_{22}=2, n_{23}=3, n_{31}=5, n_{32}=3, n_{33}=2$ bits.

Key idea: even in asymmetric cases it may be sum-capacity optimal to set $R_3=0$ when $n_{33}\leq \max\{n_{13},n_{23}\}$.
The shifted transmit signals $X_1,X_2,X_3$ arrive at the three receivers Rx$_1$, Rx$_2$, Rx$_3$ with shifts equal to $n_{ij}$, $i,j \in [1:3]$. We assume that Rx$_1$, Rx$_2$, Rx$_3$ are interested in decoding green, red and yellow bits respectively.
With $n_{33}\leq \max\{n_{13},n_{23}\},$ one might suspect that $X_3$ may convey more information to the primary receivers than  to its intended receiver. In this case, setting $R_3=0$ is optimal, and the best use of the cognitive capabilities of user~3 is to ``broadcast'' to the non-intended receivers, even in asymmetric scenarios. 
In Fig.~\ref{fig:LDA3} we give an achievable strategy for this example.
The cognitive transmitter that has in addition to its own message, message $W_1$ and $W_2$, sends a linear combination of these messages thus behaving much like a cognitive relay. Recall that the addition is bit wise over the binary field; therefore, the interference at receivers Rx$_1$ and Rx$_2$ are zero forced simultaneously (by addition modulo 2). In this example, there are no yellow bits successfully decoded at Rx$_3$ as $R_3=0$. The sum-capacity in this case is 8~bits as given in~\eqref{eq:sumrateCIFCnonsymmetric}.

\begin{figure}
\centering
\includegraphics[width=0.47\textwidth]{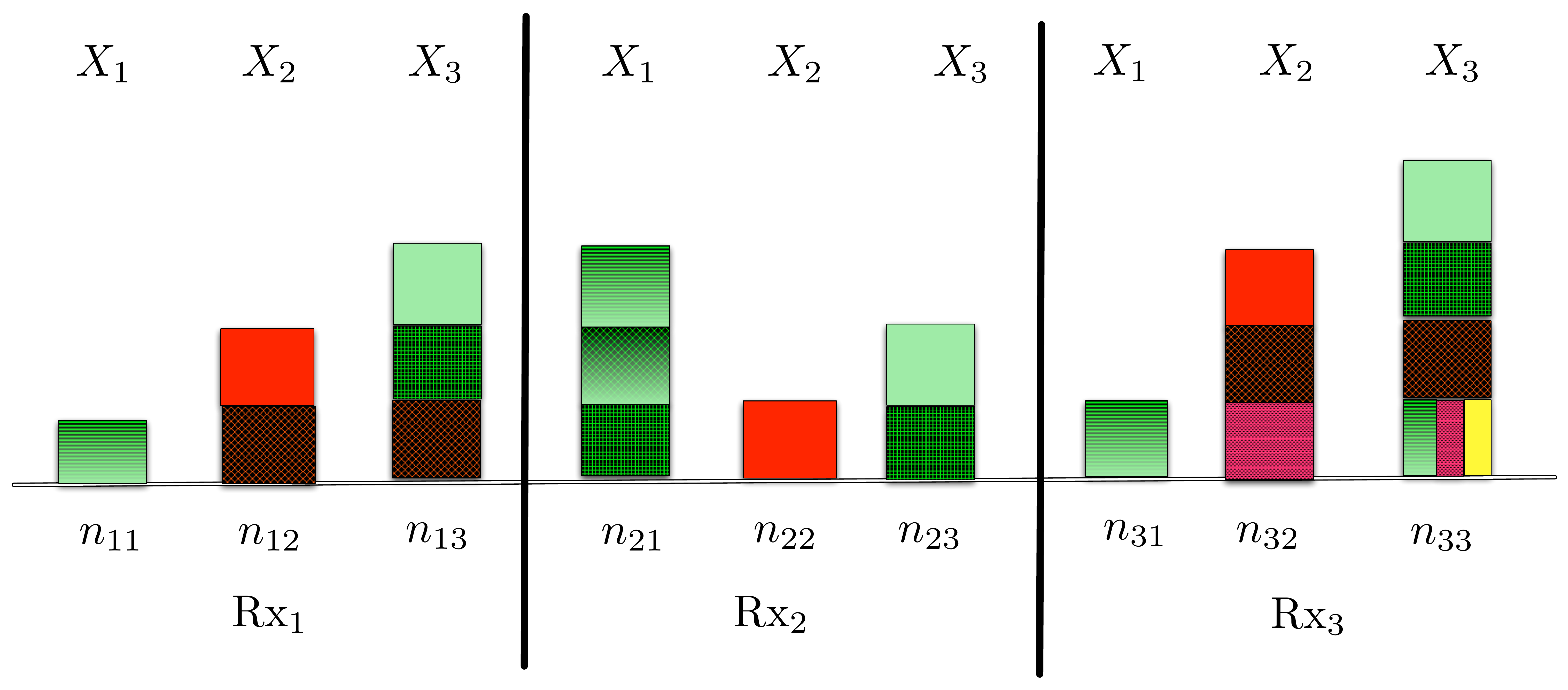}%
\caption{LDA in Example 2. The cognitive transmitter is strong enough to support a non zero rate $R_3$ and simultaneously zero force the interference at the primary receivers simultaneously. }
\label{fig:LDA2}%
\end{figure}

\paragraph{Example 2, Fig.~\ref{fig:LDA2}}\label{ex2}
Parameters: $n_{11}=1,n_{12}=2, n_{13}=3, n_{12}=1, n_{22}=3, n_{23}=2, n_{31}=1, n_{32}=3, n_{33}=4$ bits.
Key idea: when the cognitive user has $n_{33}> \max\{n_{13},n_{23}\}$, it may sneak in bits to achieve $R_3>0$, while using the cognitive relaying strategy for the other users.
The condition $n_{33}> \max\{n_{13},n_{23}\}$ suggests that the intended signal at Rx$_3$ is sufficiently strong to be able to support a non-zero rate. The form of the sum-capacity also suggests that most cognitive user can ``sneak in'' extra bits for user~3 in such a way that they do not appear at the other receivers, in other words they appear below the noise level. 
In Fig.~\ref{fig:LDA3}, schemes used in Example 1 (where user 3 acts as a cognitive relay only) are again used here, but in addition, user~3 is able to sneak in yellow bits which are below the noise level at the primary receivers by sending a combination of the interfering and desired messages, a linear deterministic version of the Gaussian dirty paper coding. The number of bits for $R_3$  in this case is $n_{33} - n_{13}$.
The sum-capacity in this case is 4~bits as given in~\eqref{eq:sumrateCIFCnonsymmetric}.

\begin{figure}
\centering
\includegraphics[width=0.47\textwidth]{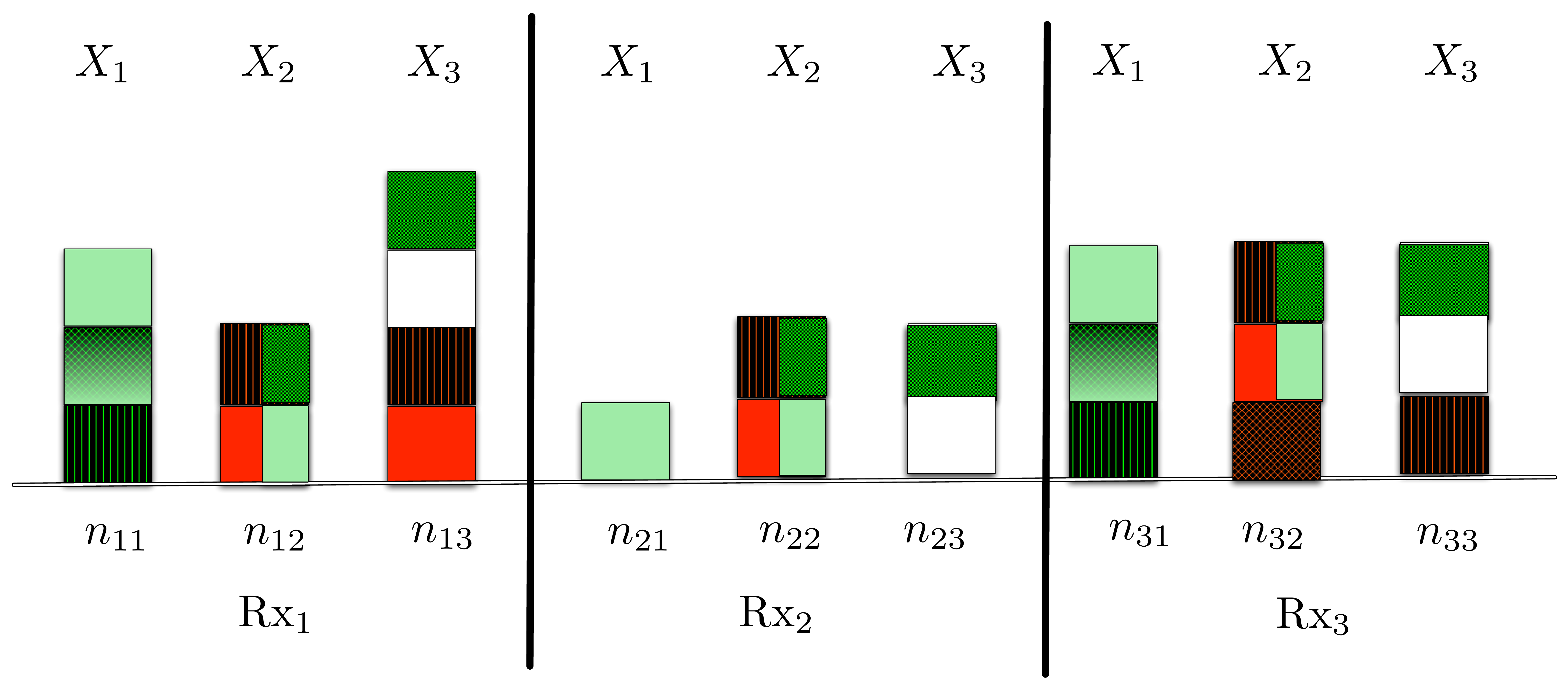}%
\caption{LDA in Example 3. The cognitive transmitter relays part of primary user~1 bits. Zero-forcing of the interference at both receivers is not an sum-capacity optimal achievability scheme. }
\label{fig:LDA1}%
\end{figure}

\paragraph{Example 3, Fig.~\ref{fig:LDA1}} 
Parameters: $n_{11}=3,n_{12}=2, n_{13}=4, n_{21}=1, n_{22}=2, n_{23}=2, n_{31}=3, n_{32}=3, n_{33}=3$ bits.
Key idea: Distributed message knowledge (not just one fully cognitive transmitter) is needed to achieve the sum-capacity outer bound (in general).
As in Example 1, we set $R_3=0$. However, notice that in this case  user~3 is not able to zero force the interference caused by primary user~1 at receiver~2.

In this case cognitive user~2 can precode against the interference caused by user~1 and thus the message knowledge structure in this case corresponds to $W_1$ at Tx$_1$, $W_1,W_2$ at Tx$_2$ and again $W_1,W_2$ at Tx$_3$. 
. 
The sum-capacity in this case is 6 bits as given in~\eqref{eq:sumrateCIFCnonsymmetric}. Note that the upper bound derived for the IFC+CR in~\cite[Theorem 2, Theorem 3, eq.(5a), eq.(8)]{DytsoCognitiverelay} when evaluated for the LDA with the channel gains of this example yields a sum-capacity of 4~bits. Therefore, message knowledge at Tx$_2$ is needed to achieve the upper bound of 6 bits as given in~\eqref{eq:sumrateCIFCnonsymmetric}.

\section{Novel Capacity Results for the AWGN}
\label{sec:NEWawgn}

The insights gained for the LDA have often been translated into Gaussian capacity results to within a constant gap for any finite SNR. In light of these success stories, we will leverage the results obtained for the $3$-user LDA in the previous section into approximate capacity results for the $3$-user Gaussian noise channel in Section~\ref{sec:NEWawgn3}. In particular, we will consider channels which satisfy channel gain relationships equivalent to those in yellow and red in Fig.~\ref{fig:shadedregion} at finite SNR. In Section~\ref{sec:NEWawgnK} we will discuss extension to the general $K$-user case.

The general Gaussian model in~\eqref{eq:AWGN channel model} 
is described by $K^2$ parameters; in the following we reduce the number of parameters involved by focusing on the {\it symmetric} case defined by: for $j\in[1:K-1],$
\begin{subequations}
\begin{align}
  &h_{jj}=|h_{\rm d}|, \ (\text{primary direct links}),
\\&h_{jK}= h_{\rm c},  \ (\text{secondary$\to$primary links}),
\\&h_{jk}= h_{\rm i},  \ k\not\in\{j,K\} \ (\text{primary interfering links}).
\end{align}
\label{eq:awgnchannelmodelsym}
\end{subequations}
The symmetric setting in~\eqref{eq:awgnchannelmodelsym} (equivalent for the $3$-user case to the parameterization in~\eqref{eq:parameterization} for the LDA) makes  
users $j\in[1:K-1]$ completely equivalent in terms of channel gains, but it does not impose any restriction on the channel gains of the cognitive receiver (i.e., $h_{Ki}, \ i\in[1:K]$) which are kept general. Moreover, the direct channel gains can be taken to be real-valued without loss of generality because the receivers 
can compensate for the phase of one of its channel gains.

\subsection{Gaussian $3$-CIFC-CMS in Strong Interference}
\label{sec:NEWawgn3}
In this section we aim to examine the relationship, in terms of sum-capacity, between the different Gaussian CIFC  models under strong interference. In particular, we will show an achievability scheme with message knowledge equivalent to that of an IFC+CR that achieves (to within a constant gap) the sum-capacity outer bound of the $3$-CIFC-CMS channel under strong interference conditions. 
Since in the LDA, a scheme with the message knowledge structure of the IFC+CR achieves an outer bound for the $3$-CIFC-CMS under strong interference conditions, highlighted in red  ($\alpha \geq 1$) in Fig.~\ref{fig:shadedregion}, intuition may suggest that a constant gap result is also possible for Gaussian noise channels at all SNR. 
In particular, we try to match the red high SNR regimes by a channel gain relationship of the form $|h_{i}|^2\geq |h_{\rm d}|^2$ for the Gaussian $3$-CIFC-CMS.

Before going into the details of the derivation, we mention that a sum-capacity achievability scheme under a strong interference condition for the LDA (see Example~2 in Fig.~\ref{fig:LDA2}, where Rx$_1$ is interfered by a strong interferer Tx$_3$ (transmit signal $X_3$ has more bits than $X_1$)) required cooperation from the strong interfererer with Tx$_1$ in the transmission of the latter's bits (green bits intended for Rx$_1$ are relayed by Tx$_3$ at the highest signal power level of the transmit signal). 
Therefore, we leverage these insights for the Gaussian channel. More precisely, we allow the cognitive transmitter to {\it coherently} beam form with the primary transmitter by allowing it to transmit the same ({\it scaled}) Gaussian transmit signal as the primary user. In the following, we give an achievability scheme for the Gaussian channel which is based on cooperation.

\begin{thm}
\label{thm:strongGaussianCMSequalsIFC+CR}
When the following holds 
\begin{subequations}
\begin{align} 
\label{eq:caseoneaKuser}
&|h_{33}|^2\leq |h_{\rm c}|^2\\ 
&\log(1+\mathbf{h}_2^H \mathbf{\Sigma}  \mathbf{h}_2) \leq \log(1+\mathbf{h}_1^H \mathbf{\Sigma}  \mathbf{h}_1)\label{eq:strongwithcovariance} 
\\&\quad   
\mathbf{h}_2^* := \begin{bmatrix} |h_{\rm d}|  \\ h_{\rm c} \\ \end{bmatrix},
\mathbf{h}_1^* := \begin{bmatrix} h_{\rm i}  \\ h_{\rm c} \\ \end{bmatrix},
\mathbf{\Sigma} :=  
\begin{bmatrix} 1-|\rho_1|^2 & \rho_3 - \rho_1 \rho_2^* \\
\rho_3^* - \rho_1^*\rho_2 & 1-|\rho_2|^2 \\ \end{bmatrix},\nonumber
\end{align}
\label{eq:conditionsstrong}
\end{subequations}
for all $|\rho_i| \leq 1, i\in[1:3]$, 
the sum-capacity outer bound of the $3$-CIFC-CMS in Theorem~\ref{thm:outer K CMS}
is achieved to within 3~bits by an achievable scheme with a message structure of IFC+CR.
\end{thm}

\begin{rem}
The regime highlighted in red in Fig.~\ref{fig:shadedregion} in the LDA is characterized by $\alpha \geq 1$, which we try to match with a channel gain relationship of the form $|h_{\rm i}|^2\geq |h_{\rm d}|^2$ for the Gaussian $3$-CIFC-CMS.
Let $|h_{\rm d}|= \mathsf{SNR}^{1/2} $, $h_{\rm i}=\mathsf{SNR}^{\alpha/2} e^{j \theta_i}$ and $h_{\rm c}=\mathsf{SNR}^{\beta/2}e^{j\theta_c}$; then, solving for~\eqref{eq:strongwithcovariance} gives 
\begin{align}
&\mathsf{SNR}(1-|\rho_1^2|)+ 2\mathsf{Re}(\rho_3-\rho_1\rho_2^*)\mathsf{SNR}^{\beta/2} \mathsf{SNR}^{1/2}e^{j\theta_c}\leq 
\nonumber
\\&
\mathsf{SNR}^{\alpha}(1-|\rho_1^2|)+ 2\mathsf{Re}(\rho_3-\rho_1\rho_2^*)\mathsf{SNR}^{\beta /2}\mathsf{SNR}^{\alpha/2} e^{j\theta_i};
\label{eq:conditionverified}
\end{align}
by taking the limit as SNR approaches infinity in~\eqref{eq:conditionverified} the condition $\max\{1,1/2+\beta/2 \}\leq \max\{\alpha,\alpha/2+\beta/2\}$ is obtained, which is equivalent to $\alpha \geq 1$.
\end{rem}

\begin{rem}
Theorem~\ref{thm:strongGaussianCMSequalsIFC+CR} shows that for the $3$-user case, full cognition only gives a constant gap improvement for the symmetric sum-capacity under strong interference conditions compared to an achievability scheme with a message structure as for the IFC+CR. 
We note however that the sum-capacity for the $K$-CIFC-CMS under strong interference conditions has been characterized in~\cite{Maamaripmsstrong} completely (for arbitrary $K$ and exact sum-capacity). The achievability scheme in that case amounts to having all cognitive users beam form to the primary receiver (as in a MISO channel). Extending our Theorem \ref{thm:strongGaussianCMSequalsIFC+CR} to arbitrary $K$, i.e., characterizing the sum-capacity to within a constant gap for the $K$-CIFC-CMS by using an achievability scheme with message knowledge of $(K-1)$-IFC+CR, is still an open problem.
\end{rem}

\begin{IEEEproof}
As a sum-capacity outer bound we use the bound initially derived for the $K$-CIFC-CMS under strong interference conditions in~\cite[Theorem 1]{Maamaripmsstrong} and proved in Appendix~\ref{strongappendix}, which specialized for the case of $K=3$ users reads:
when the following channel conditions hold
\begin{subequations}
\begin{align} 
&  I(X_{3};Y_3    |X_{1},X_2) \leq
   I(X_{3};Y_{2}|X_{1},X_2), \\
&  I(X_2,X_3;Y_2    |X_{1}) \leq
   I(X_2,X_3;Y_{1}|X_{1})
\end{align}
\label{eq:strong interf condition}
\end{subequations}
then the sum-capacity is upper bounded by
\begin{align}
R_1+R_2+R_3 \leq  I(X_1,X_2,X_3;Y_1),
\label{eq:sumcapacity}
\end{align}
where, by the `Gaussian maximizes entropy' principle~\cite[Section 2.1]{ElGamalKim:InfoTheory}, it suffices to use jointly Gaussian inputs both in~\eqref{eq:strong interf condition} and in~\eqref{eq:sumcapacity}.\footnote{
The sum-capacity upper bounds in Theorem~\ref{thm:outer K CMS} and~\eqref{eq:sumcapacity} coincide under the condition in~\eqref{eq:strong interf condition} as shown in Remark 3 in~\cite{Maamaripmsstrong}).
}
Therefore, with Gaussian inputs 
the sum-capacity upper bound in~\eqref{eq:sumcapacity} becomes
\begin{align}
\label{eq:outerboundstrongCMS}
R_1+R_2+R_3 \leq \log\Big( 1+ \big(|h_{i}|+|h_{\rm d}|+|h_{\rm c}|\big)^2\Big) =: C_{\text{sum,up}}.
\end{align}
and the channel gain conditions in~\eqref{eq:strong interf condition} reduce to~\eqref{eq:conditionsstrong},
as shown in~\cite[eq.(16) and(17)]{Maamaripmsstrong}.

For achievability, we use a scheme for the IFC+CR, where messages $W_1$ and $W_{2}$ are known at Tx$_3$. 

Tx$_1$ and Tx$_2$ use independent Gaussian random codes, and Tx$_3$ sends a superposition of the codewords generated by the primary users, and nothing for Rx$_3$. Rx$_1$ and Rx$_2$ are required to decode both messages non-uniquely. 
The achievable sum-capacity is thus as for a compound MAC and is given by
\begin{align}
\label{eq:sumcapacityinner}
&R_1+R_2+R_3 
  \leq \min_{j\in[1:2]} I(X_1,X_2,X_3; Y_j)
\\&= \log \Big( 1+\big(|h_{\rm c}|+\sqrt{|h_{\rm d}|^2+|h_{\rm i}|^2}\big)^2\Big) =: C_{\text{sum,low}}.
\label{eq:sumcapacityinnerboundstrong}\end{align}

The gap between the sum-capacity outer bound in~\eqref{eq:outerboundstrongCMS}
and the inner bound in~\eqref{eq:sumcapacityinnerboundstrong} is thus 
\begin{align*}
  &C_{\text{sum,up}}-C_{\text{sum,low}}
\\& = \log \bigg(\frac{ 1+(|h_{\rm c}|+|h_{\rm d}|+|h_{\rm i}|)^2 }{ 1+(|h_{\rm c}|+\sqrt{|h_{\rm d}|^2+|h_{\rm i}|^2})^2 
}\bigg)
\\&\stackrel{(a)}{\leq} \log \bigg(2 \left( \frac{|h_{\rm c}|+|h_{\rm d}|+|h_{\rm i}|}{|h_{\rm c}|+\sqrt{|h_{\rm d}|^2+|h_{\rm i}|^2}} \right)^2\bigg)
\\&\stackrel{(b)}{\leq} \log \bigg( 2\left(\frac{|h_{\rm c}|+2\max\{|h_{\rm d}|,|h_{\rm i}|\}}{|h_{\rm c}|+\max\{|h_{\rm d}|,|h_{\rm i}|\}}\right)^2\bigg)
\leq \log(8).
\end{align*}
where the inequalities follow form:
(a) $|h_{\rm c}|+\sqrt{|h_{\rm d}|^2+|h_{\rm i}|^2} \ge 1$, then $|h_{\rm c}|+|h_{\rm d}|+|h_{\rm i}| \ge1$ and $1+(|h_{\rm c}|+|h_{\rm d}|+|h_{\rm i}|)^2 \le 2(|h_{\rm c}|+|h_{\rm d}|+|h_{\rm i}|)^2$, and 
(b) $|h_{\rm d}|+|h_{\rm i}| \le 2\max\{|h_{\rm d}|,|h_{\rm i}|\}$ while $\sqrt{|h_{\rm d}|^2+|h_{\rm i}|^2} \ge \max\{|h_{\rm d}|,|h_{\rm i}|\}$, we see (setting $m=\max\{|h_{\rm d}|,|h_{\rm i}|\}$) that
$$
\frac{|h_{\rm c}|+|h_{\rm d}|+|h_{\rm i}|}{|h_{\rm c}|+\sqrt{|h_{\rm d}|^2+|h_{\rm i}|^2}} \le \frac{|h_{\rm c}|+2m}{|h_{\rm c}|+m} \le \frac{2|h_{\rm c}|+2m}{|h_{\rm c}|+m} =2.
$$
This completes the proof.
\end{IEEEproof}

So far we extended the LDA red region of Fig.~\ref{fig:LDA1} to the Gaussian noise case. At present, we do not have results for the LDA green region or for the asymmetric setting. The LDA yellow region of Fig.~\ref{fig:LDA1} will be discussed next.

We next show that Theorem~\ref{thm:strongGaussianCMSequalsIFC+CR}, which was inspired by Theorem~\ref{thm:IFC+CR=CoMS}, can be extended to any number of users.

\subsection{Gaussian $K$-user Cognitive Interference Channel}
\label{sec:NEWawgnK}

In this section, we present more general results for the $K$-user cognitive interference channels for $K\geq 3$ users.  Very few results exist in general for such channels. We first recall our previous result from~\cite{Maamari_JSAC} which states that a MIMO broadcast Dirty Paper Coding scheme~\cite{weingarten_MIMOBC} but with one encoding order (due to the cumulative message sharing) is sum-capacity optimal to within a constant gap for the symmetric Gaussian channel. With a particular choice of power splits (in an attempt to match the upper bound),  the message knowledge sufficient to obtain the outer bound is that  $(K-1)$-CIFC-PMS and one global cognitive user.  Such a scheme achieves the sum-capacity of the $K$-CIFC-CMS to within the following gap, despite the reduced message knowledge. 

\begin{thm}[{\cite[Theorem~4]{Maamari_JSAC}}]
\label{thm:K-user-JSAC-Gaussian}
The sum-capacity upper bound in~\eqref{eq:outer bound general K progressive} is achievable for the symmetric Gaussian $K$-CIFC-CMS to within $6$~bits 
for $K=3$ and to within $(K-2)\log(K-2) + 3.88$~bits  
for $K\geq 4$. 
\end{thm}

We now show that under specific channel gain conditions, an achievability scheme with message knowledge corresponding to that of a $K$-CIFC-CoMS and a $(K-1)$-IFC+CR achieves the $K$-CIFC-CMS outer bound (to within a constant gap). Both these results show that more message knowledge is not necessarily needed for higher gDoF (i.e., can only improve the gap).
Before going into the details of the derivation we have the following remark regarding the particular structure of the channel gain condition considered in this section.

The regime highlighted in yellow in Fig.~\ref{fig:shadedregion} in the LDA is characterized by $\alpha \leq \beta\leq 1$, which we try to match with a channel gain relationship of the form $|h_{\rm i}|^2 \leq  |h_{\rm c}|^2 \leq |h_{\rm d}|^2$ for the Gaussian $3$-CIFC-CMS. 
In particular, whenever a cognitive Tx transmits the same bits as those transmitted by the primary transmitter (see for example Example~1 in Fig.~\ref{fig:LDA3}, where $X_3$ is a combination of transmitted bits from Tx$_1$ and Tx$_2$), we match this to a zero forcing scheme for the Gaussian noise channel. Moreover, if a cognitive transmitter sneaks in bits below the noise level (see for example Example~ 2 in Fig.~\ref{fig:LDA2}, where yellow bits appeared only at Rx$_3$), we scale the Gaussian transmit signal by the interfering channel gain such that when reaches the non-intended receiver, it is received below the noise level.
In the following, we consider an equivalent $K$-user extension of the three user network and our main result is summarized in the next theorem.

\begin{thm}\label{thm:K}
When the following channel conditions hold 
\begin{align} 
\label{eq:caseoneaKuser}&|h_{KK}|\leq |h_{\rm c}|^2,  \ (K-1) |h_{\rm i}|^2 \leq  |h_{\rm c}|^2 \leq |h_{\rm d}|^2,
\end{align}
a scheme with message knowledge structure corresponding to that of a $(K-1)$-IFC+CR
achieves the sum-capacity outer bound of a $K$-CIFC-CMS to within 
\begin{align*}
&\mathsf{gap}\leq \log \left(\frac{( 2\sqrt{K-1}+K-2)^2}{(\sqrt{K-1}-1)^2}\right)+(K-1) \log(2),
\end{align*}
while when the following channel conditions hold
\begin{align} 
\label{eq:casetwoaKuser}&|h_{KK}|> |h_{\rm c}|^2, \ (K-1) |h_{\rm i}|^2 \leq  |h_{\rm c}|^2 \leq |h_{\rm d}|^2,
\end{align}
a scheme with message knowledge structure corresponding to that of a $K$-CIFC-CoMS achieves the sum-capacity outer bound of a $K$-CIFC-CMS to within 

\begin{align*}
&\mathsf{gap}\leq
\log \left( \bigg( \frac{K^2-2}{(K-1)^2}\bigg) \ \frac{(2\sqrt{K-1}+K-2)^2}{(\sqrt{K-1}-1)^2} \right) 
\\&+(K-2)\log\left(\frac{(K^2-2)}{(K-1)^2} \frac{(\sqrt{K-1}+1)^2}{(\sqrt{K-1}-1)^2}\right).
\end{align*}
\end{thm}

\begin{IEEEproof}
The sum-capacity outer bound in~\eqref{eq:outer bound general K progressive} for the channel model described in~\eqref{eq:awgnchannelmodelsym} evaluated over jointly Gaussian inputs is given by
\begin{align}\nonumber
&\sum_{k=1}^K {R_k}\leq   \log\left(1+ (|h_{\rm d}|+(K-2)|h_{\rm i}|+|h_{\rm c}|)^2  \right)\\\nonumber&+ (K-2)\log\left(1+\frac{\big||h_{\rm d}|-h_{\rm i}\big|^2}{2}\right)+ (K-2)\log(2)  \\&+ \log\left( 1+\frac{|h_{KK}|^2}{1+(K-1)|h_{\rm c}|^2}\right).
\label{sumcapacityouterboundKuser}
\end{align}

For achievability, consider the following transmit signals
\begin{subequations}
\begin{align}
& X_i = \alpha_i T_{i\text{ZF}}+ \gamma_iT_{i\text{p}}, i\in[1:K-1],
\\
& X_K = - \beta_K \sum_{i=1}^{K-1}\ T_{i\text{ZF}} + \gamma_KT_{K\text{p}},
\end{align}
\end{subequations}
where $T_{i\text{ZF}}$ and $T_{i\text{p}}$ are mutually independent $\mathcal{N}(0,1)$, for all $i\in [1:K],$ and where, for the average power constraint to be satisfy, we require 
\begin{subequations}
\begin{align}
& |\alpha_i |^2+  |\gamma_i|^2 \leq 1, \,\, i\in[1:K-1] \\
& |\gamma_K|^2 + (K-1) |\beta_K|^2 \leq 1 \,\, i\in[1:K-1].
\end{align}
\end{subequations}

For the case in~\eqref{eq:caseoneaKuser} we set 
\begin{subequations}
\begin{align}
&\gamma_i= 0,  \,\, i\in[1:K], 
\\
&\alpha_i= \beta_K = \frac{h_{\rm i}}{h_{\rm c}}.
\end{align}
\end{subequations}
With this choice of power splits and after lower bounding the achievable rate $R_1$ using the condition in~\eqref{eq:caseoneaKuser}, we have that the following rates are achievable
\begin{align}
 &R_1 =  \log \left(1+ (1-\frac{1}{\sqrt{K-1}})^2|h_{\rm d}|^2 \right)\\
 &R_i = \log \left(1+\big||h_{\rm d}|-h_{\rm i}\big|^2 \right) \,\,\, i\in[2:K-1].
\end{align}
The sum-capacity outer bound in~\eqref{sumcapacityouterboundKuser} can be upper bounded by using the condition in~\eqref{eq:caseoneaKuser} as
\begin{align}
\nonumber
&\sum_{k=1}^KR_k \leq \log\left(1+  |h_{\rm d}|^2(\frac{2\sqrt{K-1}+K-2}{\sqrt{K-1}})^2\right)  \\ 
&+(K-2) \log\left(1+\big||h_{\rm d}|-h_{\rm i} \big|^2\right)+ (K-1)\log(2).
\end{align}
The computation of the gap then follows easily.

For the case  in~\eqref{eq:casetwoaKuser}  we set  
\begin{subequations}
\begin{align}
&\gamma_i= \sqrt{\frac{1}{1+(K-1)|h_{\rm c}|^2}},  \,\, i\in[1:K], 
\\
&\alpha_i=\beta_K=\frac{h_{\rm i}}{h_{\rm c}}\sqrt{1 - \frac{1}{1+(K-1)|h_{\rm c}|^2}},
\end{align}
\end{subequations}
so that user $i\in[1:K-1]$ experience an equivalent noise power of 
\[
1+ \frac{1}{K-1}+\frac{K-2}{(K-1)^2}=\frac{K^2-2}{(K-1)^2}.
\] 
The following rates are thus achievable: for $i\in[1:K-1]$
\begin{align}
&R_i  = \log\left(1+\frac{(K-1)^2}{K^2-2}(\frac{\sqrt{K-1}-1}{\sqrt{K-1}})^2 |h_{\rm d}|^2)\right),\\
&R_{K}= \log\left(1+\frac{|h_{KK}|^2}{1+(K-1)|h_{\rm c}|^2}\right).
\end{align}
The sum-capacity outer bound in~\eqref{sumcapacityouterboundKuser} can be upper bounded by using the condition in~\eqref{eq:casetwoaKuser} as
\begin{align}
\nonumber
&\sum_{k=1}^KR_k \leq \log\left(1+  |h_{\rm d}|^2(\frac{2\sqrt{K-1}+K-2}{\sqrt{K-1}})^2\right)\\ 
\nonumber
&+(K-2)\log\left(1+  \frac{|h_{\rm d}|^2}{2}(1+\frac{1}{\sqrt{K-1}})^2\right)\\
&+ (K-2)\log(2) + \log\left( 1+\frac{|h_{33}|^2}{1+(K-1)|h_{\rm c}|^2}\right).
\end{align}
The computation of the gap then follows easily. 

We note that in both cases the gap is linear in the number of users $K$.
\end{IEEEproof}

\section{Conclusion}
\label{sec:conc}

In this paper, we surveyed and studied the relationship in terms of
capacity for different cognitive networks that differ by the amount of cognition at the transmitters. 

One general trend that emerged is that ``distributed cognition,'' or having a cumulative message knowledge structure, may not be worth the overhead as (approximately, or to within a bounded gap) the same {\it sum-capacity} can be achieved by having only one ``globally cognitive'' user whose role is to manage all the interference in the network. Whether this is true for asymmetric scenarios and for the capacity region rather than sum-capacity is an open question.
 
For multi-user cognitive networks with more than three users, many problems are still open including: characterization of the sum-capacity for asymmetric scenarios for both the linear deterministic and Gaussian networks, characterization of parameter regimes for the asymmetric Gaussian and linear deterministic Gaussian networks under which the observations made for the symmetric networks hold, and characterization of the DoF and gDoF of MIMO multi-user cognitive interference channels.


\appendices

\section{Proof of Th.~\ref{thm:IFC+CR=CoMS}}
\label{app:proof of thm IFC+CR=CoMS}

When $\alpha=1$, the sum-capacity in~\eqref{eq:sumratealphaequal1} and~\eqref{eq:sumratealpha1} are the same; therefore, we focus on the case when $\alpha\not=1$, we find the channel gain conditions that satisfy the following condition: $\eqref{eq:sumrateequation1}-\eqref{eq:secondsumrateIFC+CR}\leq 0$. We consider the six different relative orderings of $(1,\alpha,\beta)$, evaluating the bounds in~\eqref{eq:sumrateCIFCsimplfiied} and seeing under what conditions these are no larger than those in~\eqref{eq:sumrateIFC+CR}. For example if $\alpha>\beta>1$, then \eqref{eq:sumrateequation1} becomes $\alpha+\beta$,  \eqref{eq:firstsumrateIFC+CR} becomes $\alpha+\beta$ and \eqref{eq:secondsumrateIFC+CR} becomes $2\alpha+2\beta$. Hence, for this entire regime the CIFC-CMS outer bound is equal to the ICF+CR outer bound, which we know from  \cite{DytsoCognitiverelay} is achievable. Similar arguments hold for the orderings $\alpha>1>\beta$, $\beta>\alpha>1$, $\beta>1>\alpha$ and $1>\beta>\alpha$.  
In the regime $1>\alpha >\beta$
\begin{align*}
 \eqref{eq:sumrateequation1} & = 2+\beta-\alpha \\
  \eqref{eq:firstsumrateIFC+CR} & =  2+ \beta - \alpha \\
   \eqref{eq:secondsumrateIFC+CR} & = 2\max(1-\alpha,\alpha) + 2\beta.
\end{align*}
The conditions under which $2+\beta-\alpha \leq 2\max(1-\alpha,\alpha)+2\beta$ may be simplified to $2\leq 3\alpha+\beta$, which is drawn in blue in Fig.~\ref{fig:shadedregion}. In this regime, the outer bounds for the CIFC-CMS and the IFC+CR coincide. However, it is not known whether these bounds are achievable under either scheme in general. 

\section{Proof of~\eqref{eq:sumcapacity}} 
\label{strongappendix}

Let $\epsilon_N>0$ : $\epsilon_N\to 0$ as $N\to+\infty$. We have
\begin{subequations}
\begin{align*}
  &N\sum_{j=1}^{K} \left(R_j - \epsilon_N\right)
  \stackrel{\rm(a)}{\le} \sum_{j=1}^{K} I(W_j;Y^N_j) 
\\& \stackrel{\rm(b)}{\le} \sum_{j=1}^{K} I(W_j;Y^N_j | W_{[1:j-1]})
   \le \sum_{j=1}^{K} I(X^N_j;Y^N_j | X^N_{[1:j-1]})
\\& \stackrel{\rm(c)}{=}\sum_{j=1}^{K-1} I(X^N_j;Y^N_j | X^N_{[1:j-1]})+I(X^N_K;Y^N_K | X^N_{[1:K-1]})
\\& \stackrel{\rm(d)}{\le}\sum_{j=1}^{K-1} I(X^N_j;Y^N_j | X^N_{[1:j-1]})+I(X^N_K;Y^N_{K-1} | X^N_{[1:K-1]})
\\& \stackrel{\rm(e)}{=} \sum_{j=1}^{K-2} I(X^N_j;Y^N_j | X^N_{[1:j-1]})+I(X^N_{[K-1:K]};Y^N_{K-1} | X^N_{[1:K-2]})
\\& \stackrel{\rm(f)}{\le} \sum_{j=1}^{K-2} I(X^N_j;Y^N_j | X^N_{[1:j-1]})+I(X^N_{[K-1:K]};Y^N_{K-2} | X^N_{[1:K-2]})
\\&\ldots \nonumber
 \stackrel{\rm(g)}{\le} I(X^N_{[1:K]};Y^N_{1}) 
 \stackrel{\rm(h)}{\le}\sum_{t=1}^NI(X_{1t},\ldots,X_{Kt};Y_{1t}) 
\\&\ \stackrel{\rm(i)}{\le} N I(X_{[1:K]};Y_1|Q)
 \stackrel{\rm(j)}{\le} N I(X_{[1:K]};Y_1),
\end{align*}
\end{subequations}
where:
(a) follows from Fano's inequalities $H(W_j|Y^N_j) \leq N \epsilon_N, \ \forall j\in[1:K]$,
(b) from the independence of messages,
(c) by definition of encoding functions (for all $\mathcal{M}_j\subseteq [1:j]$, $j\in[1:K]$) and by the data processing inequality,
(d), (e), (f) and (g) from the condition in~\eqref{eq:strong interf condition} for $j=K$, $j=K-1$, up to $j=2$ and~\cite[Lemma 2]{Maamaripmsstrong},
(h) from the chain rule of entropy, conditioning reduces entropy and memoryless property of the channel,
(i) by introducing a time-sharing random variable $Q$ independent and uniformly distributed on $[1:N]$,
(j) by conditioning reduces entropy and since $Q$ is independent of the channel. 

\bibliography{refs}

\begin{thebibliography}{10}
\providecommand{\url}[1]{#1}
\csname url@samestyle\endcsname
\providecommand{\newblock}{\relax}
\providecommand{\bibinfo}[2]{#2}
\providecommand{\BIBentrySTDinterwordspacing}{\spaceskip=0pt\relax}
\providecommand{\BIBentryALTinterwordstretchfactor}{4}
\providecommand{\BIBentryALTinterwordspacing}{\spaceskip=\fontdimen2\font plus
\BIBentryALTinterwordstretchfactor\fontdimen3\font minus
  \fontdimen4\font\relax}
\providecommand{\BIBforeignlanguage}[2]{{%
\expandafter\ifx\csname l@#1\endcsname\relax
\typeout{** WARNING: IEEEtran.bst: No hyphenation pattern has been}%
\typeout{** loaded for the language `#1'. Using the pattern for}%
\typeout{** the default language instead.}%
\else
\language=\csname l@#1\endcsname
\fi
#2}}
\providecommand{\BIBdecl}{\relax}
\BIBdecl

\bibitem{goldsmith2009breaking}
A.~Goldsmith, S.~Jafar, I.~Maric, and S.~Srinivasa, ``Breaking spectrum
  gridlock with cognitive radios: An information theoretic perspective,''
  \emph{Proceedings of the IEEE}, vol.~97, no.~5, pp. 894--914, May 2009.

\bibitem{Shannon:1948}
C.~Shannon, ``A mathematical theory of communication,'' \emph{Bell Syst. Tech.
  J.}, vol.~27, no. 379-423, 623-656, Jul., Oct. 1948.

\bibitem{AVHESTIMERPHD}
\BIBentryALTinterwordspacing
A.~S. Avestimehr, ``Wireless network information flow: a deterministic
  approach,'' Ph.D. dissertation, EECS Department, University of California,
  Berkeley, Oct 2008. [Online]. Available:
  \url{http://www.eecs.berkeley.edu/Pubs/TechRpts/2008/EECS-2008-128.html}
\BIBentrySTDinterwordspacing

\bibitem{A_Wyner}
A.~D. Wyner, ``{Shannon-theoretic approach to a Gaussian cellular
  multiple-access channel},'' \emph{IEEE Trans. Inf. Theory}, vol.~40, no.~6,
  pp. 1713--1727, {Nov.} 1994.

\bibitem{Lozano_fun_limits}
A.~Lozano, J.~G. Andrews, and R.~W. Heath, ``{Fundamental limits of
  cooperation},'' \emph{IEEE Trans. Inf. Theory}, vol.~59, no.~9, pp.
  5213--5226, {Mar.} 2013.

\bibitem{Multicell_MIMO_Shlomo}
D.~Gesbert, S.~Hanly, H.~Huang, S.~Shamai, O.~Simeone, , and W.~Yu,
  ``{Multi-cell MIMO cooperative networks: A new look at interference},''
  \emph{IEEE J. Select. Areas Commun.}, vol.~28, no.~9, pp. 1380--1408, {Dec.}
  2010.

\bibitem{Cardone_IT}
M.~Cardone, D.~Tuninetti, R.~Knopp, and S.~Umer, ``{On the capacity of the two
  user Gaussian causal cognitive interference channel},'' \emph{IEEE Trans.
  Inf. Theory}, vol.~60, no.~5, pp. 2512--2541, {May} 2014.

\bibitem{Cardone_JSAC}
------, ``{On the Gaussian interference channel with half-duplex causal
  cognition},'' \emph{IEEE J. Select. Areas Commun.}, vol.~32, no.~11, pp.
  2177--2189, {Nov.} 2014.

\bibitem{Diana_Ergodic_TWC}
D.~Maamari, N.~Devroye, and D.~Tuninetti, ``{The sum-capacity of the ergodic
  fading Gaussian cognitive interference channel},'' \emph{IEEE Trans. Wireless
  Comm.}, vol.~14, no.~2, pp. 809--820, Sep. 2014.

\bibitem{Cover:InfoTheory}
T.~Cover and J.~Thomas, \emph{Elements of Information Theory: Second
  Edition}.\hskip 1em plus 0.5em minus 0.4em\relax Wiley, 2006.

\bibitem{ElGamalKim:InfoTheory}
{A. El Gamal and Y.-H. Kim}, \emph{Network Information Theory}.\hskip 1em plus
  0.5em minus 0.4em\relax Cambridge University Press, 2012.

\bibitem{Tse_approximate}
D.~Tse, ``{It's easier to approximate},'' in \emph{IEEE International Symposium
  on Information Theory, 2009. ISIT 2009}, Mar. 2010, pp. 6--11.

\bibitem{Maamari_JSAC}
D.~Maamari, D.~Tuninetti, and N.~Devroye, ``{Approximate sum-capacity of K-user
  cognitive interference channels with cumulative message sharing},''
  \emph{IEEE J. Select. Areas Commun.}, vol.~31, no.~11, pp. 657--667, {Mar.}
  2013.

\bibitem{Maamaripmsstrong}
------, ``{The sum-capacity of different K-user cognitive interference channels
  in strong interference},'' in \emph{Proc. IEEE Inf. Theory Workshop},
  Sevilla, Sep. 2013.

\bibitem{mahtab}
M.~Mirmohseni, B.~Akhbari, and M.~R. Aref, ``Three-user cognitive interference
  channel: Capacity region with strong interference,'' \emph{IET,
  Communications}, vol.~6, no.~13, pp. 2099--2107, {Sep.} 2012.

\bibitem{Myungstron}
M.~G. Kang and W.~Choi, ``{The capacity of a three user interference channel
  with a cognitive transmitter in strong interference},'' in \emph{Proc. IEEE
  Int. Symp. Inf. Theory}, Cambridge, Massachusetts USA, {Sep.} 2012, pp.
  1817--1821.

\bibitem{DytsoCognitiverelay}
A.~Dytso, S.~Rini, N.~Devroye, and D.~Tuninetti, ``{On the capacity region of
  the two-user interference channel with a cognitive relay},'' \emph{IEEE
  Trans. Wireless Comm.}, vol.~13, no.~12, pp. 6824--6838, {Dec.} 2014.

\bibitem{Rini:CIFC-CR}
S.~Rini, D.~Tuninetti, N.~Devroye, and A.~Goldsmith, ``{On the capacity of the
  interference channel with a cognitive relay},'' \emph{IEEE Trans. Inf.
  Theory}, vol.~60, no.~4, pp. 2148--2179, Sept. 2014.

\bibitem{etw}
R.~Etkin, D.~Tse, and H.~Wang, ``Gaussian interference channel capacity to
  within one bit,'' \emph{IEEE Trans. Inf. Theory}, vol.~54, no.~12, pp.
  5534--5562, Dec. 2008.

\bibitem{TSERATELIMITEDTXCOOP}
I.-H. Wang and D.~Tse, ``Interference mitigation through limited transmitter
  cooperation,'' \emph{Information Theory, IEEE Transactions on}, vol.~57,
  no.~5, pp. 2941--2965, May 2011.

\bibitem{devroye_IEEE}
N.~Devroye, P.~Mitran, and V.~Tarokh, ``Achievable rates in cognitive radio
  channels,'' \emph{IEEE Trans. Inf. Theory}, vol.~52, no.~5, pp. 1813--1827,
  May 2006.

\bibitem{Devroye:commag}
------, ``Limits on communications in a cognitive radio channel,''
  \emph{Communications Magazine, IEEE}, vol.~44, no.~6, pp. 44--49, June 2006.

\bibitem{Devroye:SPmag}
N.~Devroye, M.~Vu, and V.~Tarokh, ``Cognitive radio networks,'' \emph{Signal
  Processing Magazine, IEEE}, vol.~25, no.~6, pp. 12--23, November 2008.

\bibitem{gelfand}
S.~Gel'fand and M.~Pinsker, ``Coding for channels with random parameters,''
  \emph{Probl. Contr. and Inf. Theory}, vol.~9, no.~1, pp. 19--31, 1980.

\bibitem{costa}
M.~Costa, ``Writing on dirty paper,'' \emph{IEEE Trans. Inf. Theory}, vol.
  IT-29, pp. 439--441, May 1983.

\bibitem{weingarten_MIMOBC}
H.~Weingarten, Y.~Steinberg, and S.~{Shamai (Shitz)}, ``The capacity region of
  the {G}aussian {M}ultiple-{I}nput {M}ultiple-{O}utput broadcast channel,''
  \emph{IEEE Trans. Inf. Theory}, vol.~52, no.~9, pp. 3936--3964, Sep. 2006.

\bibitem{rini:journal1}
S.~Rini, D.~Tuninetti, and N.~Devroye, ``New inner and outer bounds for the
  memoryless cognitive interference channel and some new capacity results,''
  \emph{IEEE Trans. Inf. Theory}, vol.~57, no.~7, pp. 4087--4109, Jul. 2011.

\bibitem{BandemeretalAllerton2012}
B.~Bandemer, A.~{El Gamal}, and Y.-H. Kim, ``Simultaneous nonunique decoding is
  rate-optimal,'' in \emph{Proc. Allerton Conf. Commun., Control and Comp.},
  Monticello, IL, Oct. 2012.

\bibitem{motani}
H.~Chong, M.~Motani, H.~Garg, and H.~{El Gamal}, ``On the han-kobayashi region
  for the interference channel,'' \emph{IEEE Trans. Inf. Theory}, vol.~54,
  no.~7, pp. 3188--3195, Jul. 2008.

\bibitem{pramodweak}
A.~Jovicic and P.~Viswanath, ``{Cognitive Radio: an information-theoretic
  perspective},'' \emph{IEEE Trans. Inf. Theory}, vol.~55, no.~9, pp. 3945 --
  3958, Sep. 2009.

\bibitem{wu:ifcdms}
W.~Wu, S.~Vishwanath, and A.~Aripostathis, ``Capacity of a class of cognitive
  radio channels: interference channels with degraded message sets,''
  \emph{IEEE Trans. Inf. Theory}, vol.~53, no.~11, pp. 4391--4399, Nov. 2007.

\bibitem{maric_uni}
I.~Maric, R.~D. Yates, and G.~Kramer, ``The strong interference channel with
  unidirectional cooperation,'' in \emph{Proc. Workshop on Info. Theory and
  Applications}, La Jolla, Feb. 2006.

\bibitem{rini:journal2}
S.~Rini, D.~Tuninetti, and N.~Devroye, ``{Inner and outer bounds for the
  Gaussian cognitive interference channel and new capacity results },''
  \emph{IEEE Trans. Inf. Theory}, vol.~58, no.~2, pp. 820--848, Jan. 2012.

\bibitem{Sriram_MIMO_Cog}
S.~Sridharan and S.~Vishwanath, ``{On the capacity of a class of MIMO cognitive
  radios},'' \emph{IEEE J.Select. Topics in Signal Processing.}, vol.~2, no.~1,
  pp. 103--117, {Feb.} 2008.

\bibitem{rini:JSAC2014}
S.~Rini and A.~Goldsmith, ``{On the capacity of the multiantenna Gaussian
  cognitive interference channel},'' \emph{IEEE J. Select. Areas Commun.},
  vol.~32, no.~11, pp. 2252--2267, {Dec.} 2014.

\bibitem{verduwidebandslope}
S.~Verd\'u, ``{Spectral efficiency in the wideband regime},'' \emph{IEEE Trans.
  Inf. Theory}, vol.~48, no.~6, pp. 1319--1343, {Jun.} 2002.

\bibitem{tuninetti_caire_verdu:isit2002}
D.~Tuninetti, G.~Caire, and S.~Verd\'u, ``Fading multiaccess channels in the
  wideband regime: the impact of delay constraints,'' in \emph{Proceedings 2001
  {IEEE} International Symposium on Information Theory}, Lausanne, CH, June
  2002.

\bibitem{Jafar_MIMO_Cog}
C.~Huang and S.~A. Jafar, ``{Degrees of freedom of MIMO interference channel
  with cooperation and cognition},'' \emph{IEEE Trans. Inf. Theory}, vol.~55,
  no.~9, pp. 4211--4220, {Sep.} 2009.

\bibitem{nagananda2011}
K.~G. Nagananda, P.~Mohapatra, C.~R. Murthy, and S.~Kishore, ``Multiuser
  cognitive radio networks: An information theoretic perspective,'' \emph{Int.
  Journal of Adv. in Eng. Sc. and Applied Math.}, vol.~5, pp. 43--65, 2013.

\bibitem{cognitiverelaysriram}
S.~Sridharan, S.~Vishwanath, S.~A. Jafar, and S.~Shamai, ``{On the capacity of
  cognitive relay assisted Gaussian interference channel},'' in \emph{Proc.
  IEEE Int. Symp. Inf. Theory}, 2008, pp. 549--553.

\bibitem{nagananda2009information}
K.~Nagananda and C.~Murthy, ``Information theoretic results for three-user
  cognitive channels,'' in \emph{Proc. IEEE Global Telecommun. Conf.}, 2009,
  pp. 1--6.

\bibitem{nagananda2010achievable}
K.~Nagananda, C.~Murthy, and S.~Kishore, ``Achievable rates in three-user
  interference channels with one cognitive transmitter,'' in \emph{{Proc. IEEE
  Int. Sig. Proc. Commun. (SPCOM)}}, 2010, pp. 1--5.

\bibitem{mirmohseni2011capacity}
M.~Mirmohseni, B.~Akhbari, and M.~R. Aref, ``{Capacity bounds for the
  three-user cognitive Z-interference channel},'' in \emph{the 12th Canadian
  Workshop on Inf. Theory}, Kelowna, British Columbia, Canada, May 2011, pp.
  34--37.

\bibitem{vishwanathjafarianmultiusercognitive}
A.~Jafarian and S.~Vishwanath, ``On the capacity of multi-user cognitive radio
  networks,'' in \emph{Proc. IEEE Int. Symp. Inf. Theory}, 2009, pp. 601--605.

\bibitem{Erez:2004:latticecapacity}
U.Erez and R.~Zamir, ``{Achieving 1/2log( 1 + SNR) on the AWGN channel with
  lattice encoding and decoding },'' \emph{IEEE Trans. Inf. Theory}, 2004.

\bibitem{erez2005lattices}
U.~Erez, S.~Litsyn, and R.~Zamir, ``Lattices which are good for (almost)
  everything,'' \emph{IEEE Trans. Inf. Theory}, vol.~51, no.~10, pp.
  3401--3416, 2005.

\bibitem{nazer2009compute}
B.~Nazer and M.~Gastpar, ``{Compute-and-forward: Harnessing interference
  through structured codes},'' \emph{IEEE Trans. Inf. Theory}, vol.~57, no.~10,
  pp. 6463--6486, Oct. 2011.

\bibitem{Jafar:2008:alignment}
V.~Cadambe and S.~Jafar, ``Interference alignment and the degrees of freedom
  for the {K}-user interference channel,'' \emph{IEEE Trans. Inf. Theory},
  vol.~54, no.~8, pp. 3425--3441, Aug. 2008.

\bibitem{motahari2009real}
A.~Motahari, S.~Gharan, M.~Maddah-Ali, and A.~Khandani, ``{Real interference
  alignment: Exploiting the potential of single antenna systems},''
  \url{http://arxiv4.library.cornell.edu/abs/0908.2282}.

\bibitem{Jafar:2009:relays}
V.~Cadambe and S.~Jafar, ``Degrees of freedom of wireless networks with relays,
  feedback, cooperation and full duplex operation,'' \emph{IEEE Trans. Inf.
  Theory}, vol.~55, no.~5, pp. 2334--2344, May 2009.

\bibitem{Jafar:cognition}
C.~Huang and S.~A. Jafar, ``{Degrees of Freedom of the MIMO Interference
  Channel with Cooperation and Cognition},'' \emph{IEEE Trans. Inf. Theory},
  vol.~55, no.~9, pp. 4211--4220, Sep. 2008.

\bibitem{compvenu}
V.~Annapureddy, A.~{El Gamal}, and V.~Veeravalli, ``{Degrees of freedom of
  interference channels with CoMP transmission and reception },'' \emph{IEEE
  Trans. Inf. Theory}, vol.~58, no.~9, pp. 5740--5760, {Sep.} 2012.

\bibitem{Wigger_cog}
A.~Lapidoth, S.~Shamai, and M.~Wigger, ``{A linear interference network with
  local side-information },'' in \emph{Proc. IEEE Int. Symp. Inf. Theory},
  2007, pp. 2201 -- 2205.

\bibitem{Jafar_DOF_IC}
S.~A. Jafar and M.~Fakhereddin, ``{Degrees of freedom for the MIMO interference
  channel},'' \emph{IEEE Trans. Inf. Theory}, vol.~53, no.~7, pp. 2637--2642,
  {Jul.} 2007.

\bibitem{telatar_tse}
E.~Telatar and D.~Tse, ``Bounds on the capacity region of a class of
  interference channels,'' \emph{Proc. IEEE Int. Symp. Inf. Theory}, 2007.

\bibitem{suh2010feedback}
C.~Suh and D.~Tse, ``{Feedback capacity of the Gaussian interference channel to
  within 2 Bits},'' \emph{IEEE Trans. Inf. Theory}, vol.~57, no.~5, pp.
  2667--2685, May 2011.

\bibitem{arXiv:1503.07372}
M.~Cardone, D.~Tuninetti, and R.~Knopp, ``{The two-user causal cognitive
  interference channel: Novel outer bounds and constant gap result for the
  Symmetric Gaussian Noise channel in weak interference},'' \emph{Arxiv
  preprint arXiv:1503.07372}, 2015.

\end{thebibliography}
\bibliographystyle{IEEEtran}

\newpage
\begin{IEEEbiography}[{\includegraphics[width=1in,height=3.25in,clip,keepaspectratio]{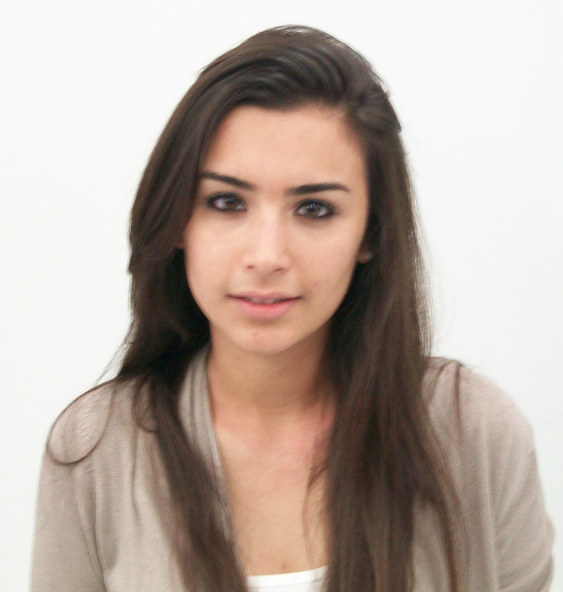}}]{Diana Maamari} is currently a research engineer at Huawei Technologies, which she joined in May 2015. Dr. Maamari received here Ph.D. in Electrical and Computer Engineering in 2015 from University of Illinois at Chicago. She received her Bachelors of Science in 2009 and Masters of Science in Electrical Engineering in 2011 from University of Balamand, Al Koura, Lebanon with high distinction. Her research during her Ph.D. focused on multi-user information theory and its applications to cognitive radio channels. Currently her research focuses on millimeter wave wireless networks.
\end{IEEEbiography}
\vspace*{-2.2cm}
\begin{IEEEbiography}[{\includegraphics[width=1in,height=1.25in,clip,keepaspectratio]{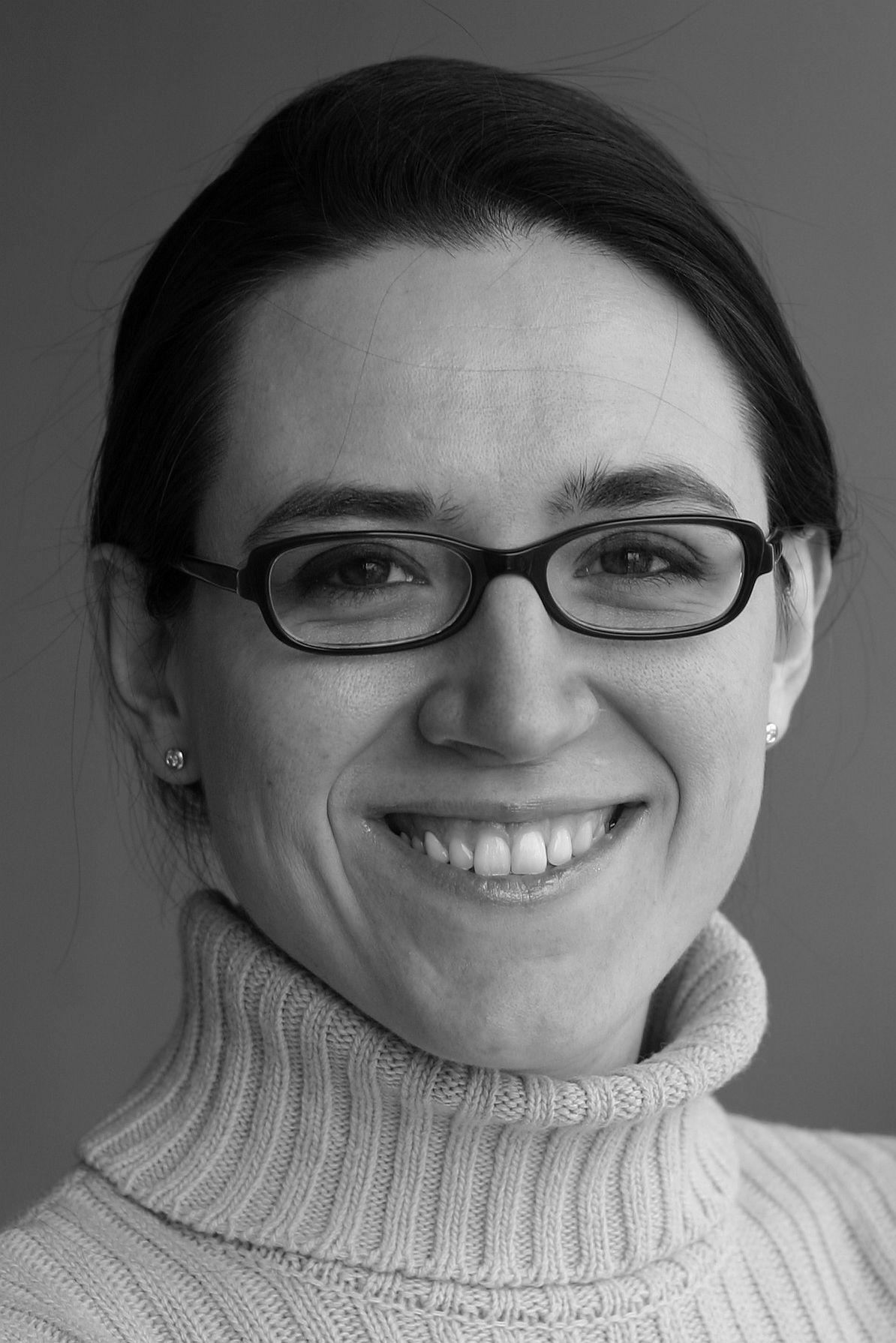}}]{Daniela Tuninetti}
is currently an Associate Professor within the Department of Electrical and Computer Engineering at the University of Illinois at Chicago (UIC), which she joined in 2005.  Dr. Tuninetti got her  Ph.D.  in  Electrical  Engineering in  2002  from  ENST/Telecom  ParisTech  (Paris,  France, with  work  done  at the  Eurecom  Institute  in  Sophia  Antipolis,  France), and   she  was  a  postdoctoral  research associate at the School of Communication and Computer Science at the Swiss Federal Institute of Technology in  Lausanne  (EPFL, Lausanne,  Switzerland) from  2002  to  2004.
Dr. Tuninetti was a recipient of a best paper award at the European Wireless Conference in 2002, of an NSF CAREER award in 2007, and named  UIC University Scholar in 2015. Dr.  Tuninetti  was  the  editor-in-chief  of  the  IEEE  Information  Theory  Society  Newsletter  from  2006  to  2008,  an  editor for IEEE  Communication  Letters  from  2006  to  2009,  and for IEEE Transactions on Wireless Communications  from  2011  to  2014; she is currently an associate editor for IEEE Transactions on Information Theory.  Dr. Tuninetti's research interests are in the ultimate performance limits of wireless interference networks, with special emphasis on cognition and user cooperation, in coexistence between radar and communication systems, in multi-relay networks, and in content-type coding.
\end{IEEEbiography}
\vspace*{-2.2cm}
\begin{IEEEbiography}[{\includegraphics[width=1in,height=1.25in,clip,keepaspectratio]{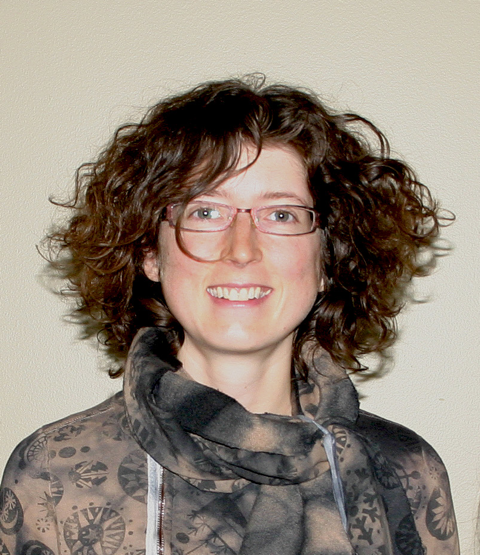}}]{Natasha Devroye}
is an Associate Professor in the Department of Electrical and Computer Engineering at the University of Illinois at Chicago (UIC), which she joined in January 2009. From July 2007 until July 2008 she was a Lecturer at Harvard University. Dr. Devroye obtained her Ph.D in Engineering Sciences from the School of Engineering and Applied Sciences at Harvard University in 2007,  and a Honors B. Eng in Electrical Engineering from McGill University in 2001.   Dr. Devroye was a recipient of an NSF CAREER award in 2011 and was named UIC's Researcher of the Year in the ``Rising Star'' category in 2012. She has been an Associate Editor for IEEE Transactions on Wireless Communications, IEEE Journal of Selected Areas in Communications, and is currently an Associate Editor for the IEEE Transactions on Cognitive Communications and Networking. Her research focuses on multi-user information theory and applications to cognitive and software-defined radio, radar, relay and two-way communication networks.
\end{IEEEbiography}

\end{document}